# EEA Professional Climate Survey Report

Tim Lee, Massimo Morelli, Marvin Pappalettera, Dario Sansone, Sulin Sardoschau[*]

**Executive Summary**

The European Economic Association (EEA) Minorities in Economics (MinE) Committee, together with the German Economic Association (VfS), conducted a professional climate survey in 2023. The survey targeted current and former members of the EEA with a focus on equality, diversity, and inclusiveness, particularly across the dimensions of gender, ethnicity, LGBTQ+ status, and disability. The aim was to assess the professional climate for minorities within the economics profession across Europe. The effort is comparable to the American Economic Association (AEA) professional climate surveys, against which we also compare our results below.

*Demographic breakdown*

The total number of respondents for the survey was 861. The demographic breakdown of survey respondents highlights diversity across gender, ethnicity, LGBTQ+ identity, and disability. Women comprised 46.5% of the participants. Ethnic minorities accounted for 22.6% of the sample, including Chinese, Asian, Black, Muslim, and Sinti/Roma individuals. LGBTQ+ representation was 17.3%, with 9.8% identifying as non-heterosexual, transgender, or non-binary. Additionally, 25% of respondents reported having a physical condition that affected their work or studies.

Geographically, there was a concentration in German-speaking countries, with Germany leading at 35%, followed by the UK (12%), Nordic countries (10%), Italy (7%), and France (5%). Eastern Europe had lower representation, with only 15 respondents. Note that this is where the respondents hold their current job, which is not necessarily where they are from.

Demographic variation across regions was significant. In Spain and Portugal, 28.6% of respondents identified as ethnic minorities. The Netherlands and Belgium had the highest LGBTQ+ representation at 26.4%. Eastern Europe had the highest proportion of respondents with disabilities at 40%, but the fewest ethnic minorities at 6.7%. Respondents from Eastern Europe were generally older, with fewer participants under 40 years old.

We also asked respondents about their socioeconomic background. The majority of respondents grew up in a household with high socioeconomic status (65.4%) and raised by parents with tertiary education (59.2%). France in particular had a notably low share of individuals with parents without a university degree compared to other European countries.

---

[*] Sang Yoon (Tim) Lee, Queen Mary University of London, TSE and CEPR, sylee.tim@qmul.ac.uk; Massimo Morelli, Bocconi University, LISER and CEPR, massimo.morelli@unibocconi.it; Marvin Pappalettera, Bocconi University, marvin.pappalettera@studbocconi.it; Dario Sansone, University of Exeter and IZA, d.sansone@exeter.ac.uk; Sulin Sardoschau, Humboldt University, sulin.sardoschau@hu-berlin.de. We thank Jan Eeckhout, Gilat Levy, Gemma Prunner-Thomas and Vincent Sterk for contributing throughout various stages while preparing the survey and the report.

## Perceptions about the professional climate in economics in Europe

Significant disparities were found across demographic lines. Women, ethnic minorities, and LGBTQ+ respondents reported less satisfaction with the professional climate compared to their counterparts. Only 15% of women, compared to 31% of men, expressed satisfaction with the overall professional climate in economics. Social and intellectual inclusion also varied widely, with only 22% of women and 23% of LGBTQ+ respondents reporting that they felt socially included, compared to 40% of men.

There was a marked gender gap in perceptions of respect within the field. While 86% of respondents believed men are respected in economics, only 33% believed the same for women. LGBTQ+ respondents also felt less respected, with only 20% of respondents reporting respect for transgender and gender non-conforming individuals in the field. While the overall number was a bit higher for people with disabilities, only 18% of women reported that people with disabilities are respected in the profession - as opposed to 46% of men.

## Discrimination, exclusion, and harassment

Discrimination was a prominent issue, particularly among ethnic minorities and LGBTQ+ individuals. Over 30% of ethnic minorities and 26% of LGBTQ+ respondents reported experiencing discrimination within the field, compared to 19% of men. Women were significantly more likely to report sex-based discrimination, with nearly half (49%) citing such experiences. In addition, it is worth noting that 33% of respondents witnessed discrimination or unfair treatment based on sex, and 25% based on marital status or caregiving responsibilities. Women also reported a higher incidence of specific discriminatory practices. For instance, 32% reported hostility during seminars (16% for men), and 28% felt they were treated unfairly in student evaluations (5% for men). 32% of women reported they often avoided disclosing or discussing their marital status or caregiving responsibilities (10% for men).

In addition, women, ethnic minorities, LGBTQ+ individuals, and people with disabilities often took actions to avoid possible discrimination and harassment, such as not applying for certain jobs or accepting admission to certain graduate programs, quitting their job, or not attending social events. In particular, LGBTQ+ individuals and people with disabilities were twice as likely to avoid talking about their ideas, and women and people with disabilities were disproportionately unwilling to participate in discussions. Likewise, women and people with disabilities reported the highest rate of feeling excluded.

Harassment was also high among minorities. Experiencing inappropriate behavior was reported by 42% of women and 36% of people with disabilities. Women often received unwanted romantic or sexual attention (20%). 3% of women, 3% of ethnic minorities 3% of LGBTQ+ respondents reported having been sexually assaulted by another economist.

Finally, women and people with disabilities who are also from lower socioeconomic backgrounds reported even higher rates of discrimination.

Geographic Differences

The EEA survey reveals significant regional differences in the professional climate across Europe. The sample average for the overall climate was 3.85 on a 6-point scale. Respondents in Nordic countries (Denmark, Finland, Iceland, Norway, and Sweden) reported the highest satisfaction with a climate score of 4.18 and one of the lowest incidences of sexual harassment (5%) and discrimination. Italy scored among the lowest, with an average climate score of 3.67, with 56.3% reporting having experienced discrimination. Respondents in German-speaking countries (Germany, Switzerland, Austria) also showed dissatisfaction, with 55.9% reporting discrimination. Those in the UK and Ireland reported the highest rate of discrimination at 57.4%. Conversely, respondents in Spain and Portugal reported the lowest rate of discrimination (45.9%) but the highest rate of having experienced sexual harassment (11.1%), reflecting a mixed professional environment.

The comparison between the EEA 2023 and AEA 2018 surveys shows that EEA respondents reported lower satisfaction with the professional climate. Only 23% of EEA respondents were satisfied, compared to 34% in the AEA, with EEA women reporting even lower satisfaction than their AEA counterparts (15% vs. 20%). Feelings of being valued were also lower in the EEA, with 28% of all respondents feeling valued compared to 40% in the AEA; For women, the numbers were 21% vs. 25%. Social inclusion was also lower among all EEA respondents (31% in the EEA vs. 37% in the AEA). Intellectual inclusion followed this trend, with 37% of EEA respondents feeling included compared to 42% in the AEA.

Ethnic minorities in the EEA reported the lowest satisfaction, at 19%, compared to 34% for non-whites in the AEA. LGBTQ+ and disabled respondents in the EEA also felt less valued, with only 26% of LGBTQ+ and 20% of disabled respondents reporting positive experiences, compared to 34% and 30%, respectively, in the AEA.

Challenges and Limitations

The survey, conducted in collaboration with the German Economic Association, reached out to current and past EEA members and was promoted through various channels to enhance participation. Despite these efforts, the response rate was significantly lower than comparable surveys, such as the AEA's 2018 survey. In particular, the share of respondents from a high SES background is rather high at 65.4%, and as many as 25% of respondents reported having a disability (compared to 10% in the AEA 2018 survey), both of which could indicate over-representation of such groups. Conversely, the EEA sample has much less ethnic minorities compared to the AEA, which could indicate a smaller presence in Europe, but also under-representation.

The small sample size also limited our ability to condition the responses on multiple characteristics due to privacy concerns. For example, women and people with disabilities who are also from lower socioeconomic backgrounds reported even higher rates of discrimination, suggesting we could learn more from digging deeper into more finely-defined demographic categories.

While acknowledging the importance of the experiences reported in the survey, the low response rate limits the representativeness of the findings.

**Introduction and Background**

In October 2023, the EEA Minorities in Economics Committee (MinE), supported by the EEA Executive Committee and in cooperation with the German Economic Association ("Verein für Socialpolitik," VfS), implemented a professional climate survey. The survey was sent to all current members of the EEA (as of January 2024), as well as all individuals who had been a member in the preceding ten years and still signed up for EEA newsletters. MinE also solicited support from department heads and graduate program directors of all higher education institutions known to the EEA. The survey was then advertised through several channels, such as the European Committee for LGBTQ+ Economists. In addition, the EEA created and shared through social media a video with several economists emphasizing the importance of completing the survey.

Survey questions focused on individuals' experiences and perceptions of equality, diversity, and inclusiveness, especially along the dimensions of gender, race and ethnicity, LGBTQ+ status, and disability.[1] Care was taken to ensure that the survey would capture differences in the treatment (or perceived treatment) of individuals across the different countries in Europe. While the survey itself was designed, organized, and monitored by MinE, the actual implementation and storage of data was administered by Dynata, a third-party operator with no prior association with the EEA, to ensure data privacy and anonymity.

This report summarizes MinE's work before, during, and after the survey, including an analysis and interpretation of the results. It is no secret that the survey was modeled after the AEA's professional climate surveys in 2018 and 2023. However, despite the AEA including numerous members not necessarily based in North America, the EEA decided that a separate survey is necessary to capture its unique context where the professional network spans numerous countries with significant differences in their institutions (as compared to across American states) and the relatively more recent influx of non-Europeans into the area. Moreover, the EEA survey added more emphasis on the experiences of doctoral students in graduate programs, as well as early career researchers.

The rest of the report is organized as follows. Section 1 outlines the survey methodology, response rate, and a summary of our interactions with Dynata while designing the survey and collecting results. Section 1.2, in particular, discusses the types of selection and/or response bias we might expect, especially given that the response rate was rather low.

Section 2 presents the main results of interest in the form of tables. We begin with a broad overview of key results, followed by tables that summarize the shares of positive responses to some of our main survey questions, such as respondents' perception of the professional climate, direct or indirect experiences of discrimination, and measures respondents state that they have taken to avoid unfair treatment, exclusion or harassment. The results are also cross-tabulated by gender, whether or not the respondent is (or perceives themselves to be) an ethnic minority, LGBTQ+ status, and disability. For each question, we highlight the key takeaways by each of these demographic dimensions. We also present results from two regressions that can potentially capture composition effects.

---

[1] At the EEA, gender issues would fall under the remit of the Women in Economics Committee (WinE) rather than MinE, but MinE took the lead on the survey. Other dimensions such as family background and religion were also considered.

Section 3 disaggregates the responses by country (or groups of countries). By doing so, we not only highlight the differences in the profession's status across countries/regions but can also identify where and along which dimensions the EEA community should make more efforts to improve the status of minorities in the economic profession.

Section 4 compares our results against other surveys. The comparison with the AEA's 2018 climate survey reveals similarities and differences in the challenges faced by the economics profession on the two sides of the Atlantic, while the comparisons against broader European surveys sheds light on whether our professional environment faces issues that coincide with, or diverge from, the general European population. The European surveys we consider are "Discrimination in the European Union" from 2023, which is part of the special modules of the Eurobarometer (link), and the "Labour Force Survey" (EU-LFS) from 2021 – Eurostat (link).

**1. Survey Design, Data Collection and Storage**

The implementation of the survey, data collection, and storage was entirely delegated to Dynata (https://www.dynata.com/). At no point in our analysis did anyone from the EEA, VfS, or MinE directly observe or handle the raw data gathered from the survey. Dynata was specifically chosen after contacting several data platforms because they have abundant experience working with European firms while having minimal overlap with PhD economists, thus minimizing the possibility of data privacy being compromised.

Starting from the AEA's 2018 Professional Climate Survey as a template, MinE modified and expanded the questionnaire to adapt to the European context. Coordination with Dynata began and continued throughout the summer of 2023. A link to a self-administered web survey went live on January 12, 2024, and the EEA sent invitations to current and past (dating back ten years) EEA members. We also asked department heads and graduate program directors of all higher education institutions known to the EEA to disseminate the link within their respective departments, including their doctoral students. The survey was closed on April 4, 2024.

The total number of respondents for the survey was 861. This is rather small compared to AEA's 2018 survey, which gathered 10,406 responses (22.9% response rate). Table 1 summarizes the respondents' characteristics.

Women represent 46.5% of the sample, compared to 30% in the AEA 2018 survey. One of the survey questions asked respondents whether they consider themselves part of one or more cultural, ethnic, or religious groups. Respondents who identified as Chinese, any Asian, Muslim, Sinti/Roma, Black, or "None of the above" (including "Prefer Not To Answer" or PNTA) were categorized as Minority, comprising 22.6% of the sample.[2]

The LGBTQ+ variable was constructed from a survey question that asked respondents to select the option that best described their sexual orientation. A value of 1 was assigned to individuals selecting Gay, Lesbian, Bisexual, "Other", "Don't Know", or PNTA, and a value of 0 to those identifying as Heterosexual. Overall, 149 respondents (17.3%) identified as LGBTQ+. More specifically, 84 individuals (9.8%) identified as nonheterosexual, transgender/non-binary, or both, and 7 individuals (0.8%) identified as transgender or non-binary (or both).

Another survey question asked respondents whether they had a physical condition that affected their work or studies, with approximately 25% of respondents answering "Yes" or selecting PNTA.

Regarding the "Academic capital" variable, individuals with at least one parent holding a master's degree or above or whose parents both have a bachelor's degree were coded as 1, while everyone else was coded as 0. Around 40.8% of respondents fell into the latter category. On the other hand, respondents' socio-economic status (SES) is measured based on two separate questions. The first question asked how the respondent's household managed to cope with the monthly income when they were about 14 years old: "very well", "fairly well", "somewhat well", "somewhat poorly", "fairly poorly", and "very poorly". The second question asked respondents how they would classify the socio-economic status of the household in which they grew up: "high income/wealthy", "upper middle class", "middle class", "low income/working class", and "in poverty". Anyone who answered "very well" or "fairly well" in the first question or "high income/wealthy" in the second was coded as "High SES" and all the others as low. 563 respondents fell in the high socio-economic status category (65.4%).

The "Perceived as Ethnic Minority" variable was based on a question asking respondents whether, irrespective of the cultural, ethnic, or religious groups they belong to, others perceive them as part of one or more such groups. Respondents who selected the option Chinese, any Asian, Muslim, Sinti/Roma, Black, or "None of the above" (including PNTA) were coded as 1, while the ones selecting Jewish, Eastern European, or White (and not 1) were coded as 0. In total, 158 individuals (18.4%) in the sample were perceived as ethnic minorities.

The "Year of Doctoral Degree" variable was coded as 1 if the respondent received their doctoral degree from 2010 onward and 0 otherwise, with 419 individuals (48.7%) falling into this category.

Overall, the average age of respondents is around 39, and 72.5% are married, with a slightly lower marriage rate among women at 69.3%. Moreover, around 78% of those employed work for a university and 12.1% work for another kind of research institution. Regarding political leaning, around 38.7% of respondents are politically left-wing, 51% centrist, 4.6% right-wing, and 5.7% prefer not to answer. Finally, around 7.4% of all respondents are students.

---

[2] Unlike other demographic variables, respondents were allowed to choose multiple categories for this question, thus a strictly exclusive categorization was not possible. 6% of respondents responded PNTA.

Table 2 and Figure 1 summarize response rates across the European continent. It is evident that the joint organization of the survey with the VfS resulted in a rather large share of responses originating from Germany and other German-speaking countries overall (302 and 65, respectively)[3]. Other countries with a large share of respondents were the United Kingdom (107), Italy (63), and France (41).

Table 3 shows that labor economists represent the largest share of respondents (286 respondents), followed by microeconomists (190), and then public economists (160).

## 1.1 Data Delivery

MinE was in close contact with Dynata throughout the data collection period and after, allowing MinE to monitor progress. The survey was initially to be closed on March 10, 2024, but was extended to April 2024 due to low response rates.

Dynata then released the results to MinE in two stages. The first stage was pre-contracted, containing coefficients from two multivariate regressions and conditional means to a limited set of survey questions. Once the number of respondents and results from the first stage were known, MinE asked for additional results consisting of additional conditional means by demographic. No conditional means were released for groups consisting less than 4.5% of the sample.

## 1.2 Sample Response and Selection Bias

The EEA sample may not be representative of the underlying population in the economics profession in Europe. Respondents may be selected along socio-demographic dimensions or may hold different views with respect to the overall professional climate. Since there is no comprehensive data set on the composition of European academics in economics at different career stages, we are not able to benchmark sample selection based on observables. However, we can compare the characteristics of the EEA professional climate sample with those of other surveys run by the EEA and other economics associations. In addition, we also focus on differences in respondent characteristics and outcomes across European countries.

Table 18 compares respondent characteristics between the EEA and the AEA round 1 climate survey. Both respondent groups show a high percentage of employment in colleges or universities (78% for EEA, 75% for AEA). A significant difference is observed in gender representation, with EEA having 46.5% female respondents compared to AEA's 30%. Context-specific aspects of measuring and conceptualizing minority status in Europe and North America are mirrored in the respective surveys. The EEA has a lower percentage of Asian (5.8% vs. 15%) and Black respondents (0.7% vs. 3%) than AEA. Conversely, 23% of EEA respondents identify with an ethnic minority, while only 18% are perceived as an ethnic minority. Additionally, EEA respondents report higher LGBTQ+ representation (17.3% vs. 6%)[4] and disability prevalence (25% vs. 10%). EEA respondents are also generally younger (mean age 39.5 vs. 47.3). Both groups share a high rate of marriage among women but differ in dependency responsibilities, with EEA women having fewer dependents (42.3% vs. 49%). Importantly, these differences need not point to sample bias in either survey but instead reflect differences in the underlying populations.

---

[3] Switzerland and Luxembourg were included as "German-speaking" countries due to their academic environment being close to Germany's, but more importantly because categorizing them as their own separate group(s) could threaten respondents' anonymity. Dynata was asked to release results for specific groups only if they exceeded 4.5% of the total sample.

Lastly, survey response rates varied substantially across countries of employment. As the EEA survey was heavily promoted by the VfS partner, the largest group of respondents (360 in total and 41.8% of all survey respondents) are employed in Germany, Switzerland and Austria at the time of the survey. The second and third largest groups of respondents come from the UK and Italy. Table 2 reports summary statistics by regional group.

There are significant differences across country groups, as we explain in detail in Section 2. Some notable results are that the Netherlands and Belgium recorded a high share of LGBTQ+ respondents (26.4%), and Eastern Europe a high share of respondents with disabilities (40%). At the same time, respondents in Eastern Europe were older, with only 26.7% below the age of 40, and the least likely to belong to an ethnic minority (6.7%). Interestingly, while Spain and Portugal have the lowest share of respondents reporting having experienced discrimination at 45.9%, they also reported the highest share of having experienced sexual harassment (11.1%). Nordic countries were estimated to have the highest satisfaction with the general climate and the lowest incidence of sexual harassment.

The EEA sample's representativeness of the broader European economics profession is uncertain, and it is crucial to interpret the results with caution. The socio-demographic selection of respondents and the potential for varying views on the professional climate may influence the findings. However, despite these limitations, the survey provides valuable insights into the characteristics and perceptions of a significant segment of the economics profession. Indeed, even if not representative, these responses clearly show that a large number of economists do not feel welcome or valued and that numerous individuals faced discrimination and harassment. The selection process in the survey may not allow us to provide population-based estimates, but the data does support the claim that discrimination exists and is widespread in the profession and that this is not an issue specific to one country alone. The comparisons with the AEA survey further highlight meaningful differences and similarities that can inform future research and policy discussions.

---

[4] LGBTQ+ includes PNTA (prefer not to answer). Excluding PNTA, EEA respondents still report an LGBTQ+ representation of 10.1 percent.

Table 1: Respondents' characteristics.

|  | Number of respondents | Mean |
|---|---|---|
| Female | 400 | 0.465 |
| Cultural group: |  |  |
| White | 659 | 0.765 |
| Black | 6 | 0.007 |
| Eastern European | 35 | 0.041 |
| Jewish | 15 | 0.017 |
| Ethnic minority | 195 | 0.226 |
| Perceived as Ethnic Minority | 158 | 0.184 |
| LGBTQ+ | 149 | 0.173 |
| Transgender and/or non-binary | 7 | 0.008 |
| With a disability | 215 | 0.250 |
| Student | 64 | 0.074 |
| Academic capital | 510 | 0.592 |
| Not in high socio-economic group | 298 | 0.346 |
| Doctoral degree from 2010 onwards | 419 | 0.487 |
| Among employed, employer is: |  |  |
| College or University | 672 | 0.780 |
| For-profit organization | 4 | 0.005 |
| Not-for-profit organization | 4 | 0.005 |
| Federal government or EU institution | 29 | 0.034 |
| Non-University Research institution | 104 | 0.121 |
| Age | 861 | 39.446 |
| Married | 624 | 0.725 |
| Politically left-wing | 333 | 0.387 |
| Among women: |  |  |
| Married | 277 | 0.693 |
| With dependents | 169 | 0.423 |
| Employer is a college or university | 308 | 0.770 |

Table 2: Response rates by region.

|  | Number of respondents | Mean |
|---|---|---|
| Germany | 302 | 0.35 |
| UK | 107 | 0.12 |
| Ireland | 8 | 0.01 |
| Nordic | 83 | 0.10 |
|     Denmark | 31 | 0.04 |
|     Finland | 7 | 0.01 |
|     Iceland | 1 | 0.00 |
|     Norway | 19 | 0.02 |
|     Sweden | 25 | 0.03 |
| German speaking | 65 | 0.08 |
|     Austria | 31 | 0.04 |
|     Switzerland | 27 | 0.03 |
|     Luxembourg | 7 | 0.01 |
| Italy | 63 | 0.07 |
| Iberia | 42 | 0.05 |
|     Spain | 36 | 0.04 |
|     Portugal | 5 | 0.01 |
|     Andorra | 1 | 0.00 |
| France | 41 | 0.05 |
| Netherlands | 38 | 0.04 |
| Belgium | 15 | 0.02 |
| Eastern Europe | 15 | 0.02 |
| PNTA | 57 | 0.07 |
| The rest (Greece, Australia, US, Canada, Israel, Turkey, UAE, China, Japan, Mexico) | 25 | 0.03 |

Table 3: Respondents by economics sub-field.

| | Number of respondents | Mean |
|---|---|---|
| General Econ and Teaching/History of Econ Thought, Method and Heterodox | 34 | 0.039 |
| Law and Economics/Business Admin, Business Economics, marketing, Accounting, etc. | 50 | 0.058 |
| Mathematical and Quantitative Methods | 60 | 0.070 |
| Microeconomics | 190 | 0.221 |
| Macroeconomics and Monetary Economics | 150 | 0.174 |
| International Economics | 97 | 0.113 |
| Financial Economics | 67 | 0.078 |
| Public Economics | 160 | 0.186 |
| Health, Education, and Welfare | 138 | 0.160 |
| Labor and Demographic Economics | 286 | 0.332 |
| Industrial Organization | 69 | 0.080 |
| Economic History | 41 | 0.048 |
| Economic Development, Innovation, Technological Change, and Growth | 128 | 0.149 |
| Political Economy and Comparative Economic Systems | 79 | 0.092 |
| Agricultural and Natural Resource Economics / Environmental and Ecological Economics | 79 | 0.092 |
| Urban, Rural, Regional, Real Estate, and Transportation Economics | 49 | 0.057 |
| Misc Categories/PNTA | 71 | 0.082 |
| Other Special Topics/JEL codes do not fit my research | 34 | 0.039 |

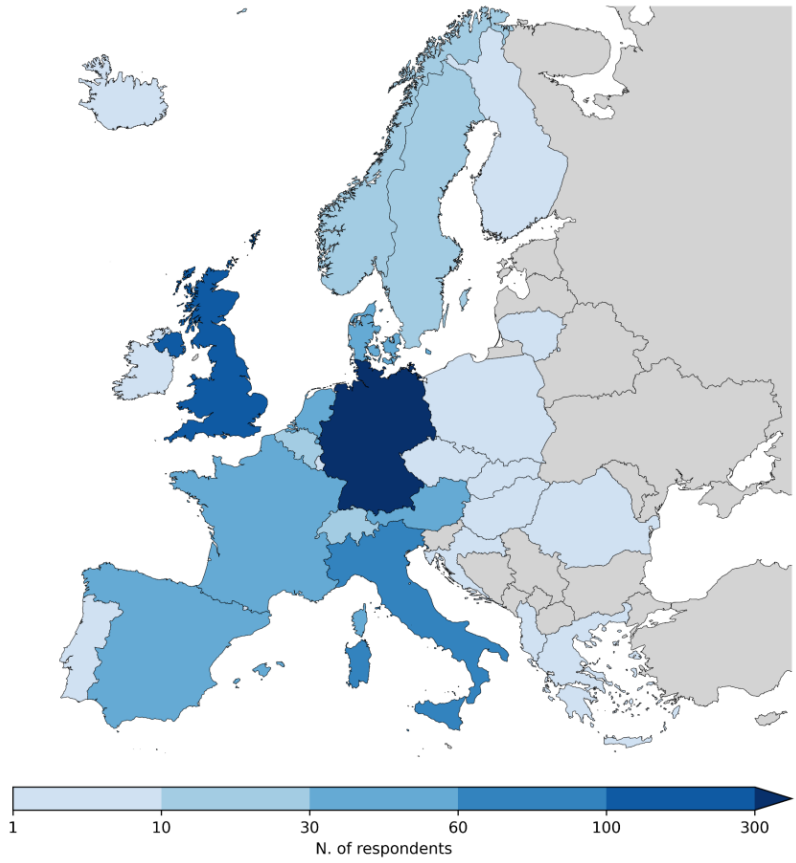

Figure 1: Number of responses across the European continent.

## 2. Main results

### 2.1 General Climate

Table 4 summarizes the results of the general climate questions. The share of respondents who *agree* or *strongly agree* with each statement is reported in each cell. Each question was presented on a 6-point scale; other possible answers were: "strongly disagree", "disagree", "somewhat disagree", and "somewhat agree". The results are reported for the overall population as well as by gender, LGBTQ+ status, disability status, and ethnic minorities (Table 4A). In addition, we also report the results for individuals with no academic capital, those perceived as ethnic minorities, the ones not in high socio-economic status, and those who received their doctoral degree from 2010 onwards (Table 4B).

In general, men consistently report more positive experiences compared to other groups. For example, men are twice as likely as women to be satisfied with the overall climate in the field of economics: 31% of men agree or strongly agree with the statement "I am satisfied with the overall professional climate within the field of economics" compared to only 15% of women. This gender gap is also evident in satisfaction with the general climate at the institutional level (63% of men vs. 45% of women). A similar disparity is also present between LGBTQ+ and non-LGBTQ+ respondents, with the former group reporting lower satisfaction with the general climate at the institutional level (45%) compared to the latter (55%). Individuals with a disability also tend to be less satisfied with the climate at their institution compared to individuals without a disability (40% vs. 58%). Individuals who are perceived as ethnic minorities and those who are not in high socio-economic status also tend to report lower satisfaction with the general climate at their institution (43% and 45%).

The discrepancies among these groups are also present in other areas of the general climate. When it comes to feeling valued, men again report the highest levels both within the field (36%) and at their institutions (62%). In contrast, women, those with disabilities, LGBTQ+ respondents, those perceived as ethnic minorities, and the ones who are not in high socio-economic status feel less valued.

Another area of concern is social and intellectual inclusion, particularly for women, LGBTQ+ individuals, those with a disability, and ethnic minorities. While 40% of men always feel socially included within the field, only 22% of women and 23% of LGBTQ+ respondents feel the same. This is also true for intellectual inclusion, where just 29% of women and 27% of respondents with a disability report always feeling included intellectually within the field, compared to 46% of men. These patterns are also present at the institutional level, where men again report the highest levels of inclusion, and women, LGBTQ+ individuals, those with a disability, and ethnic minorities report the lowest.

Another significant issue highlighted by the data is discrimination. The highest levels of discrimination within the field of economics are reported by individuals who are perceived as ethnic minorities (34%), those who belong to an ethnic minority (31%), and LGBTQ+ (26%). On the other hand, only 19% of men report experiencing discrimination. At the institutional level, discrimination is reported less frequently, but again, there are significant differences between female and male respondents (16% vs. 8%); moreover, respondents with a disability report higher levels of discrimination compared to other groups (21%).

Respondents with disabilities also agree the least with the statement "the work that I do is valued at my institution/place of employment" (43%), while men agree the most (60%). Again, there are significant differences between LGBTQ+ respondents and non-LGBTQ+ respondents (46% vs. 56%) and between respondents with a disability and those without (43 vs. 58%).

Finally, ethnic minorities are more likely to feel that their ideas and opinions are often ignored within the field of economics compared to other respondents (22% vs. 16% overall).

## 2.2 Opinions and Perceptions

Table 5 summarizes the results from different questions related to opinions and perceptions. The share of respondents who agree or strongly agree with each statement is reported in each cell. Other possible answers were "strongly disagree", "disagree", "somewhat disagree", and "somewhat agree".

A significant share of respondents (66%) believe that economics would be a more vibrant discipline if it were more diverse. This belief is firm among women (83%) and is the weakest among men (50%).

Only 34% of respondents perceived as part of an ethnic minority, and 42% of ethnic minorities believe that people of their ethnicity are respected within the field of economics, compared to 74% of respondents overall.

Regarding gender, 86% of respondents believe that men are respected within the field, with almost all women sharing this view (96%). However, only 33% of respondents believe that women are respected, with this figure dropping to 13% among women themselves, indicating a sharp contrast in perceptions of gender respect within the field.

Women and LGBTQ+ respondents perceive lower levels of respect for transgender and gender non-conforming people (both around 13%) and for people who are not heterosexual (33% and 29%) compared to men (26% and 58%, respectively). Finally, only 18% of women, compared to 46% of men, believe that people with a disability are respected within the field of economics.

## 2.3 Experiences of Discrimination

Table 6 reports the experiences of discrimination or unfair treatment based on various personal attributes. In each cell, the share of respondents who report having personally experienced such discrimination or unfair treatment by anyone in the field of economics over the last ten years in the field of economics in Europe is shown.

Similarly, Table 7 reports the share of respondents who have *witnessed* discrimination in the same period in the field of economics in Europe.

The survey highlights a significant issue for women, with a considerable share reporting discrimination due to marital status or caregiving responsibilities. This finding aligns with existing research on the "child penalty" in academia, which explores the impact of parenthood on academic careers ("Parenthood and academic career trajectories", A. S. Lassen and R. Ivandić., 2024). Women are significantly more likely to face sex-based discrimination compared to men, with 49% of women reporting such experiences versus 12% of men. Conversely, 38% of men, compared to 27% of women, report having witnessed discrimination based on sex.

Moreover, individuals perceived as ethnic minorities, as well as those who identify as ethnic minorities, are much more likely to report discrimination based on their racial or ethnic identity (27% and 20%, respectively, vs. 7% overall) and their citizenship status (24% and 19% vs. 10% overall).

### 2.4 Experiences of discrimination outside of academia

Table 8 reports the experiences of discrimination or unfair treatment based on various personal attributes. In each cell is shown the share of respondents that report having personally experienced such discrimination or unfair treatment by anyone in the field of economics in their current job.

Similarly, Table 9 reports the share of respondents who have *witnessed* discrimination in the same period in their current job.

The results reveal a pattern similar to that observed in the field of economics. Women are once again significantly more likely to experience sex-based discrimination compared to men (32% vs. 6%). Additionally, individuals perceived as ethnic minorities, as well as those who identify as ethnic minorities, are much more likely to report discrimination based on their racial or ethnic identity (18% and 13%, respectively, compared to 6% overall) and their citizenship status (16% and 13%, compared to 7% overall).

### 2.5 Undisclosed information

Table 10 reports personal attributes respondents may have avoided disclosing or discussing due to fear of negative consequences, harassment, or discrimination. Each cell reports the share of respondents who report having personally avoided disclosing or discussing such attributes over the last ten years in the field of economics in Europe.

Overall, political views are the most avoided topic, with 39% of respondents indicating they have avoided discussing them. This trend is consistent across gender, with 38% of men and women choosing to withhold their political views. Individuals with disabilities are the most likely to avoid discussing their political views (52%). Moreover, LGBTQ+ respondents show a slightly higher tendency to avoid this discussion (48%) compared to non-LGBTQ+ respondents (37%).

When it comes to marital status and caregiving responsibilities, there is an evident disparity between men and women, with 32% of women reporting they have avoided discussing such topics compared to just 10% of men.

Sexual orientation is particularly avoided by LGBTQ+ respondents, with 50% indicating they have avoided disclosing or discussing this aspect, a much more significant percentage compared to the 13% overall average.

Finally, respondents with disability and LGBTQ+ are more likely to refrain from discussing research topics (30% and 33%) compared to respondents without disabilities and non-LGBTQ+ (both around 18%).

### 2.6 Experiences as a student

Table 11 shows experiences of discrimination or unfair treatment as a student. The share of respondents who answered "yes" to each question is reported in each cell.

Notably, women are more likely than men (21% vs. 10%) to report being discriminated against in terms of access to quality advising.

### 2.7 Experiences of discrimination in academia

Table 12 reports experiences of discrimination or unfair treatment in various aspects of work. In each cell, the share of respondents who felt that they had personally experienced discrimination or unfair treatment in such an aspect by anyone in the last ten years in the field of economics in Europe is reported.

The results underscore significant gender disparities in experiences of discrimination in academia. Women are notably more likely than men to report facing discrimination across several key areas. Specifically, women are more likely to report discrimination in terms of promotion decisions (24% vs. 13%), compensation (27% vs. 12%), teaching assignments (21% vs. 8%), service obligations (29% vs. 7%), membership in a committee or decision-making body (24% vs. 10%), course evaluations (28% vs. 5%), and access to potential coauthors (18% vs. 5%). Women are also more likely than men to face hostility during presentations and talks (32% vs. 16%).

### 2.8 Avoidance

Table 13 reports the results relative to the actions respondents may have taken to avoid possible harassment, discrimination, or unfair or disrespectful treatment by one or more economists. Each cell reports the share of respondents who report having taken the listed action over the last ten years in the field of economics in Europe.

Overall, the results reveal that it is common for respondents to avoid presenting a question, idea, or view at their place of work, with 40% of respondents indicating so. This tendency is particularly pronounced among women (54%), while it is the least evident among men (26%). Such gender disparity is also present in other areas. For example, women are more likely than men to report not speaking at a conference or during a seminar presentation (56% vs. 30%), not attending social events (43% vs. 17%), and not starting or continuing research in a particular field (28% vs. 18%).

Similar disparities are also evident between respondents with and without disabilities. Those with disabilities are more likely to avoid presenting ideas at work (48% vs. 37%), speaking at conferences or seminars (55% vs. 38%), attending social events (41% vs. 26%), and initiating or continuing research in particular fields (33% vs. 20%).

### 2.9 Exclusion and Harassment

Table 14 reports on experiences of exclusion and harassment. Each cell shows the share of respondents who report having personally experienced such treatment over the last ten years in the field of economics in Europe.

Table 15 provides additional information on experiences of assault.

Overall, 59% of respondents reported feeling socially excluded at a meeting or event, with this experience being more prevalent among women (69%) compared to men (50%). Respondents with disabilities (68%) and individuals not in high socio-economic status (66%) also tend to report higher rates compared to others.

Feeling disrespected by economist colleagues is also a shared experience, with 56% of respondents reporting it. This sentiment is again expressed more frequently by women (69%) and individuals with disabilities (62%), compared to 44% of men and 54% of those without disabilities.

Around 65% of respondents also reported feeling their work was not taken as seriously as that of their economist colleagues. The highest rates are reported by women (75%) and respondents with disabilities (68%) compared to 57% of men. Additionally, 56% of respondents (64% of women vs. 50% of men) feel that the subject or methodology of their research is not taken as seriously as that of their economist colleagues.

In terms of inappropriate behavior, 42% of women have experienced offensive remarks or gestures from other economists or students, and around 20% have reported unwanted attempts by other economists or students to establish a romantic or sexual relationship. These percentages are much lower among men, at 10% and 3%, respectively.

Though less common, severe incidents of harassment and assault are reported, with 3% of women having experienced sexual assault and 7% reporting being stalked by another economist or student.

**2.10 Multivariate analysis**

Tables 16 and 17 report linear regression analyses for the variables related to the general climate and experiences of exclusion and harassment.

In general, LGBTQ+ is almost never significant, and ethnic minorities are also almost never significant in the harassment part (while more so in the agree-to-disagree part). On the other hand, both the female as well as the disability dummies are very strong in both parts of the survey. Moreover, low socio-economic status seems to be acting as an amplification factor.

Table 4A: General climate.

| | All | Female | Male | LGBTQ+ | Not LGBTQ+ | With disability | No disability | Ethnic minority |
|---|---|---|---|---|---|---|---|---|
| I am satisfied with the overall professional climate within the field of economics. | 0.23 | 0.15 | 0.31 | 0.21 | 0.23 | 0.21 | 0.24 | 0.19 |
| I am satisfied with the overall professional climate at my institution/place of employment. | 0.54 | 0.45 | 0.63 | 0.45 | 0.55 | 0.40 | 0.58 | 0.46 |
| I feel valued within the field of economics. | 0.28 | 0.21 | 0.36 | 0.26 | 0.29 | 0.20 | 0.31 | 0.22 |
| I feel valued at my institution/place of employment. | 0.54 | 0.47 | 0.62 | 0.46 | 0.56 | 0.40 | 0.59 | 0.48 |
| I always feel included socially within the field of economics. | 0.31 | 0.22 | 0.40 | 0.23 | 0.33 | 0.24 | 0.33 | 0.22 |
| I always feel included socially at my institution/place of employment. | 0.57 | 0.49 | 0.67 | 0.49 | 0.59 | 0.50 | 0.60 | 0.50 |
| I always feel included intellectually within the field of economics. | 0.37 | 0.29 | 0.46 | 0.34 | 0.38 | 0.27 | 0.40 | 0.33 |
| I always feel included intellectually at my institution/place of employment. | 0.55 | 0.46 | 0.63 | 0.47 | 0.56 | 0.42 | 0.59 | 0.51 |
| I feel I have been discriminated against within the field of economics. | 0.22 | 0.25 | 0.19 | 0.26 | 0.21 | 0.27 | 0.20 | 0.31 |
| I feel I have been discriminated against at my institution/place of employment. | 0.13 | 0.16 | 0.08 | 0.14 | 0.12 | 0.21 | 0.10 | 0.17 |
| The work that I do is valued within the field of economics. | 0.36 | 0.32 | 0.39 | 0.29 | 0.37 | 0.29 | 0.38 | 0.34 |
| The work that I do is valued at my institution/place of employment. | 0.54 | 0.49 | 0.60 | 0.46 | 0.56 | 0.43 | 0.58 | 0.51 |
| I think I have a great deal of power within the field of economics. | 0.04 | 0.03 | 0.05 | 0.03 | 0.04 | 0.02 | 0.05 | 0.05 |
| I think I have a great deal of power at my institution/place of employment. | 0.12 | 0.09 | 0.15 | 0.07 | 0.13 | 0.07 | 0.13 | 0.10 |
| My ideas and opinions are often ignored within the field of economics. | 0.16 | 0.14 | 0.16 | 0.21 | 0.14 | 0.21 | 0.14 | 0.22 |
| My ideas and opinions are often ignored at my institution/place of employment. | 0.12 | 0.13 | 0.10 | 0.13 | 0.11 | 0.15 | 0.11 | 0.17 |

Reported in each cell is the share of respondents who agree or strongly agree with each statement.

Table 4B: General climate.

| | No academic capital | Perceived as ethnic minority | Doctoral degree from 2010 onwards | Not in high SES |
|---|---|---|---|---|
| I am satisfied with the overall professional climate within the field of economics. | 0.23 | 0.21 | 0.21 | 0.20 |
| I am satisfied with the overall professional climate at my institution/place of employment. | 0.50 | 0.43 | 0.56 | 0.45 |
| I feel valued within the field of economics. | 0.26 | 0.22 | 0.26 | 0.24 |
| I feel valued at my institution/place of employment. | 0.51 | 0.46 | 0.54 | 0.48 |
| I always feel included socially within the field of economics. | 0.29 | 0.26 | 0.27 | 0.26 |
| I always feel included socially at my institution/place of employment. | 0.54 | 0.46 | 0.58 | 0.49 |
| I always feel included intellectually within the field of economics. | 0.35 | 0.32 | 0.35 | 0.37 |
| I always feel included intellectually at my institution/place of employment. | 0.51 | 0.50 | 0.56 | 0.46 |
| I feel I have been discriminated against within the field of economics. | 0.25 | 0.34 | 0.25 | 0.26 |
| I feel I have been discriminated against at my institution/place of employment. | 0.15 | 0.18 | 0.11 | 0.17 |
| The work that I do is valued within the field of economics. | 0.32 | 0.34 | 0.33 | 0.31 |
| The work that I do is valued at my institution/place of employment. | 0.51 | 0.49 | 0.57 | 0.48 |
| I think I have a great deal of power within the field of economics. | 0.03 | 0.04 | 0.02 | 0.04 |
| I think I have a great deal of power at my institution/place of employment. | 0.11 | 0.12 | 0.11 | 0.11 |
| My ideas and opinions are often ignored within the field of economics. | 0.18 | 0.20 | 0.17 | 0.20 |
| My ideas and opinions are often ignored at my institution/place of employment. | 0.14 | 0.13 | 0.09 | 0.17 |

Reported in each cell is the share of respondents who agree or strongly agree with each statement.

Table 5A: Opinions and perceptions.

| | All | Female | Male | LGBTQ+ | Not LGBTQ+ | With disability | No disability | Ethnic minority |
|---|---|---|---|---|---|---|---|---|
| It is not important for the field of economics to be inclusive towards people with different backgrounds. | 0.11 | 0.09 | 0.12 | 0.13 | 0.11 | 0.12 | 0.11 | 0.14 |
| Discrimination is rare within the field of economics. | 0.06 | 0.03 | 0.08 | 0.04 | 0.06 | 0.06 | 0.06 | 0.05 |
| Economics would be a more vibrant discipline if it were more diverse. | 0.66 | 0.83 | 0.50 | 0.68 | 0.65 | 0.67 | 0.65 | 0.65 |
| People of my ethnicity are respected within the field of economics | 0.74 | 0.75 | 0.76 | 0.63 | 0.77 | 0.67 | 0.77 | 0.42 |
| People who are not White are respected within the field of economics | 0.36 | 0.24 | 0.48 | 0.29 | 0.38 | 0.32 | 0.38 | 0.31 |
| Men are respected within the field of economics | 0.86 | 0.96 | 0.79 | 0.81 | 0.88 | 0.84 | 0.87 | 0.83 |
| Women are respected within the field of economics | 0.33 | 0.13 | 0.51 | 0.30 | 0.34 | 0.32 | 0.33 | 0.31 |
| Transgender and gender non-conforming people are respected within the field of economics | 0.20 | 0.13 | 0.26 | 0.13 | 0.21 | 0.21 | 0.20 | 0.21 |
| People of my sexual orientation are respected within the field of economics | 0.72 | 0.67 | 0.78 | 0.41 | 0.78 | 0.60 | 0.75 | 0.59 |
| People who are not heterosexual are respected within the field of economics | 0.46 | 0.33 | 0.58 | 0.29 | 0.49 | 0.42 | 0.47 | 0.39 |
| People with a disability are respected within the field of economics | 0.33 | 0.18 | 0.46 | 0.27 | 0.34 | 0.28 | 0.34 | 0.29 |
| At your university, the discussion culture during academic presentations is productive and fair | 0.54 | 0.45 | 0.63 | 0.51 | 0.54 | 0.47 | 0.56 | 0.49 |
| At international conferences in Europe, the discussion culture during academic presentations is productive and fair | 0.41 | 0.35 | 0.48 | 0.33 | 0.43 | 0.38 | 0.43 | 0.34 |

Reported in each cell is the share of respondents who agree or strongly agree with each statement.

Table 5B: Opinions and perceptions.

| | No academic capital | Perceived as ethnic minority | Doctoral degree from 2010 onwards | Not in high SES |
|---|---|---|---|---|
| It is not important for the field of economics to be inclusive towards people with different backgrounds. | 0.11 | 0.15 | 0.11 | 0.15 |
| Discrimination is rare within the field of economics. | 0.06 | 0.05 | 0.05 | 0.05 |
| Economics would be a more vibrant discipline if it were more diverse. | 0.60 | 0.66 | 0.67 | 0.59 |
| People of my ethnicity are respected within the field of economics | 0.74 | 0.34 | 0.72 | 0.66 |
| People who are not White are respected within the field of economics | 0.39 | 0.28 | 0.35 | 0.35 |
| Men are respected within the field of economics | 0.85 | 0.80 | 0.88 | 0.78 |
| Women are respected within the field of economics | 0.36 | 0.28 | 0.30 | 0.33 |
| Transgender and gender non-conforming people are respected within the field of economics | 0.19 | 0.23 | 0.19 | 0.19 |
| People of my sexual orientation are respected within the field of economics | 0.72 | 0.57 | 0.73 | 0.64 |
| People who are not heterosexual are respected within the field of economics | 0.44 | 0.40 | 0.45 | 0.43 |
| People with a disability are respected within the field of economics | 0.36 | 0.27 | 0.32 | 0.34 |
| At your university, the discussion culture during academic presentations is productive and fair | 0.53 | 0.45 | 0.58 | 0.49 |
| At international conferences in Europe, the discussion culture during academic presentations is productive and fair | 0.44 | 0.37 | 0.42 | 0.40 |

Reported in each cell is the share of respondents who agree or strongly agree with each statement.

Table 6A: Directly experienced discrimination in the field of economics - by type of discrimination.

| | All | Female | Male | LGBTQ+ | Not LGBTQ+ | With disability | No disability | Ethnic minority |
|---|---|---|---|---|---|---|---|---|
| Have you ever been discriminated against, or treated unfairly based on: | | | | | | | | |
| Racial/ethnic identity | 0.07 | 0.07 | 0.07 | 0.09 | 0.06 | 0.08 | 0.07 | 0.20 |
| Sex | 0.30 | 0.49 | 0.12 | 0.31 | 0.29 | 0.34 | 0.28 | 0.27 |
| Gender identity | 0.07 | 0.11 | 0.04 | 0.08 | 0.07 | 0.08 | 0.07 | 0.08 |
| Sexual orientation | 0.02 | 0.01 | 0.02 | 0.07 | 0.01 | 0.02 | 0.02 | 0.03 |
| Disability status | 0.02 | 0.02 | 0.01 | 0.03 | 0.01 | 0.06 | 0.00 | 0.03 |
| Marital status/caregiving responsibilities | 0.13 | 0.20 | 0.06 | 0.10 | 0.13 | 0.14 | 0.12 | 0.14 |
| Religion | 0.02 | 0.01 | 0.03 | 0.01 | 0.02 | 0.02 | 0.02 | 0.04 |
| Political views | 0.09 | 0.08 | 0.09 | 0.12 | 0.09 | 0.11 | 0.09 | 0.14 |
| Age | 0.14 | 0.18 | 0.10 | 0.13 | 0.14 | 0.15 | 0.13 | 0.16 |
| Citizenship status | 0.10 | 0.11 | 0.09 | 0.15 | 0.09 | 0.14 | 0.08 | 0.19 |
| Place of employment | 0.23 | 0.27 | 0.20 | 0.23 | 0.24 | 0.26 | 0.23 | 0.27 |
| Research topics | 0.27 | 0.33 | 0.20 | 0.32 | 0.25 | 0.30 | 0.25 | 0.33 |

Reported in each cell is the share of respondents who report having personally experienced different types of discrimination or unfair treatment over the last ten years in the field of economics in Europe.

Table 6B: Directly experienced discrimination in the field of economics - by type of discrimination.

|  | No academic capital | Perceived as ethnic minority | Doctoral degree from 2010 onwards | Not in high SES |
|---|---|---|---|---|
| Have you ever been discriminated against, or treated unfairly based on: | | | | |
| Racial/ethnic identity | 0.07 | 0.27 | 0.08 | 0.10 |
| Sex | 0.29 | 0.26 | 0.31 | 0.30 |
| Gender identity | 0.05 | 0.06 | 0.07 | 0.07 |
| Sexual orientation | 0.02 | 0.03 | 0.02 | 0.02 |
| Disability status | 0.02 | 0.01 | 0.01 | 0.03 |
| Marital status/caregiving responsibilities | 0.13 | 0.12 | 0.12 | 0.14 |
| Religion | 0.02 | 0.04 | 0.02 | 0.03 |
| Political views | 0.11 | 0.15 | 0.06 | 0.12 |
| Age | 0.13 | 0.16 | 0.13 | 0.14 |
| Citizenship status | 0.10 | 0.24 | 0.11 | 0.13 |
| Place of employment | 0.25 | 0.28 | 0.27 | 0.29 |
| Research topics | 0.30 | 0.36 | 0.27 | 0.32 |

Reported in each cell is the share of respondents who report having personally experienced different types of discrimination or unfair treatment over the last ten years in the field of economics in Europe.

Table 7A: Witnessed discrimination in the field of economics - by type of discrimination.

|  | All | Female | Male | LGBTQ+ | Not LGBTQ+ | With disability | No disability | Ethnic minority |
|---|---|---|---|---|---|---|---|---|
| Have you ever witnessed discrimination or unfair treatment based on: | | | | | | | | |
| Racial/ethnic identity | 0.21 | 0.27 | 0.15 | 0.26 | 0.20 | 0.27 | 0.19 | 0.21 |
| Sex | 0.33 | 0.27 | 0.38 | 0.34 | 0.32 | 0.29 | 0.34 | 0.32 |
| Gender identity | 0.15 | 0.18 | 0.11 | 0.24 | 0.13 | 0.22 | 0.12 | 0.16 |
| Sexual orientation | 0.08 | 0.11 | 0.06 | 0.14 | 0.07 | 0.09 | 0.08 | 0.08 |
| Disability status | 0.06 | 0.07 | 0.05 | 0.09 | 0.05 | 0.09 | 0.05 | 0.06 |
| Marital status/caregiving responsibilities | 0.25 | 0.33 | 0.18 | 0.31 | 0.24 | 0.30 | 0.24 | 0.23 |
| Religion | 0.07 | 0.07 | 0.06 | 0.08 | 0.06 | 0.11 | 0.05 | 0.08 |
| Political views | 0.20 | 0.20 | 0.22 | 0.19 | 0.21 | 0.26 | 0.19 | 0.19 |
| Age | 0.21 | 0.24 | 0.17 | 0.23 | 0.20 | 0.20 | 0.21 | 0.19 |
| Citizenship status | 0.18 | 0.21 | 0.15 | 0.16 | 0.18 | 0.17 | 0.18 | 0.17 |
| Place of employment | 0.29 | 0.29 | 0.29 | 0.27 | 0.29 | 0.28 | 0.29 | 0.28 |
| Research topics | 0.32 | 0.33 | 0.29 | 0.32 | 0.32 | 0.34 | 0.31 | 0.27 |

Reported in each cell is the share of respondents who have witnessed different types of discrimination or unfair treatment over the last ten years in the field of economics in Europe.

Table 7B: Witnessed discrimination in the field of economics - by type of discrimination.

|  | No academic capital | Perceived as ethnic minority | Doctoral degree from 2010 onwards | Not in high SES |
|---|---|---|---|---|
| Have you ever witnessed discrimination or unfair treatment based on: | | | | |
| Racial/ethnic identity | 0.21 | 0.23 | 0.22 | 0.24 |
| Sex | 0.28 | 0.33 | 0.34 | 0.29 |
| Gender identity | 0.12 | 0.19 | 0.14 | 0.15 |
| Sexual orientation | 0.08 | 0.11 | 0.09 | 0.08 |
| Disability status | 0.05 | 0.08 | 0.06 | 0.06 |
| Marital status/caregiving responsibilities | 0.23 | 0.22 | 0.28 | 0.21 |
| Religion | 0.06 | 0.11 | 0.07 | 0.09 |
| Political views | 0.21 | 0.20 | 0.19 | 0.20 |
| Age | 0.18 | 0.22 | 0.21 | 0.22 |
| Citizenship status | 0.17 | 0.19 | 0.17 | 0.16 |
| Place of employment | 0.25 | 0.27 | 0.31 | 0.26 |
| Research topics | 0.30 | 0.25 | 0.33 | 0.32 |

Reported in each cell is the share of respondents who have witnessed different types of discrimination or unfair treatment over the last ten years in the field of economics in Europe.

Table 8A: Directly experienced discrimination in the current job - by type of discrimination.

|  | All | Female | Male | LGBTQ+ | Not LGBTQ+ | With disability | No disability | Ethnic minority |
|---|---|---|---|---|---|---|---|---|
| Have you ever been discriminated against, or treated unfairly based on: | | | | | | | | |
| Racial/ethnic identity | 0.06 | 0.07 | 0.04 | 0.06 | 0.06 | 0.06 | 0.06 | 0.13 |
| Sex | 0.18 | 0.32 | 0.06 | 0.15 | 0.19 | 0.23 | 0.17 | 0.14 |
| Gender identity | 0.04 | 0.06 | 0.02 | 0.03 | 0.04 | 0.06 | 0.04 | 0.03 |
| Sexual orientation | 0.02 | 0.01 | 0.02 | 0.03 | 0.01 | 0.02 | 0.01 | 0.00 |
| Disability status | 0.01 | 0.02 | 0.01 | 0.03 | 0.01 | 0.04 | 0.00 | 0.03 |
| Marital status/caregiving responsibilities | 0.09 | 0.15 | 0.04 | 0.07 | 0.09 | 0.12 | 0.08 | 0.09 |
| Religion | 0.02 | 0.01 | 0.03 | 0.01 | 0.02 | 0.01 | 0.02 | 0.03 |
| Political views | 0.07 | 0.07 | 0.07 | 0.09 | 0.07 | 0.08 | 0.07 | 0.11 |
| Age | 0.10 | 0.14 | 0.07 | 0.08 | 0.10 | 0.11 | 0.09 | 0.10 |
| Citizenship status | 0.07 | 0.08 | 0.06 | 0.09 | 0.06 | 0.12 | 0.05 | 0.13 |
| Place of employment | 0.10 | 0.11 | 0.10 | 0.09 | 0.11 | 0.12 | 0.10 | 0.11 |
| Research topics | 0.17 | 0.21 | 0.14 | 0.20 | 0.16 | 0.20 | 0.16 | 0.20 |

Reported in each cell is the share of respondents who report having personally experienced different types of discrimination or unfair treatment by anyone in the field of economics in their current job.

Table 8B: Directly experienced discrimination in the current job - by type of discrimination.

| | No academic capital | Perceived as ethnic minority | Doctoral degree from 2010 onwards | Not in high SES |
|---|---|---|---|---|
| Have you ever been discriminated against, or treated unfairly based on: | | | | |
| Racial/ethnic identity | 0.05 | 0.18 | 0.05 | 0.08 |
| Sex | 0.20 | 0.15 | 0.18 | 0.20 |
| Gender identity | 0.03 | 0.02 | 0.04 | 0.04 |
| Sexual orientation | 0.02 | 0.00 | 0.02 | 0.01 |
| Disability status | 0.02 | 0.01 | 0.01 | 0.02 |
| Marital status/caregiving responsibilities | 0.09 | 0.08 | 0.07 | 0.10 |
| Religion | 0.01 | 0.03 | 0.02 | 0.02 |
| Political views | 0.08 | 0.10 | 0.06 | 0.09 |
| Age | 0.07 | 0.11 | 0.09 | 0.11 |
| Citizenship status | 0.06 | 0.16 | 0.07 | 0.09 |
| Place of employment | 0.09 | 0.11 | 0.10 | 0.11 |
| Research topics | 0.18 | 0.23 | 0.14 | 0.22 |

Reported in each cell is the share of respondents who report having personally experienced different types of discrimination or unfair treatment by anyone in the field of economics in their current job.

Table 9A: Witnessed discrimination in the current job - by type of discrimination.

| | All | Female | Male | LGBTQ+ | Not LGBTQ+ | With disability | No disability | Ethnic minority |
|---|---|---|---|---|---|---|---|---|
| Have you ever witnessed discrimination or unfair treatment based on: | | | | | | | | |
| Racial/ethnic identity | 0.12 | 0.13 | 0.10 | 0.15 | 0.11 | 0.18 | 0.09 | 0.13 |
| Sex | 0.23 | 0.20 | 0.25 | 0.24 | 0.22 | 0.22 | 0.23 | 0.26 |
| Gender identity | 0.09 | 0.10 | 0.08 | 0.17 | 0.07 | 0.12 | 0.08 | 0.11 |
| Sexual orientation | 0.05 | 0.06 | 0.05 | 0.09 | 0.05 | 0.07 | 0.05 | 0.05 |
| Disability status | 0.03 | 0.04 | 0.02 | 0.05 | 0.03 | 0.06 | 0.02 | 0.04 |
| Marital status/caregiving responsibilities | 0.15 | 0.20 | 0.10 | 0.20 | 0.14 | 0.17 | 0.15 | 0.15 |
| Religion | 0.04 | 0.05 | 0.03 | 0.04 | 0.04 | 0.06 | 0.04 | 0.08 |
| Political views | 0.11 | 0.11 | 0.10 | 0.14 | 0.11 | 0.19 | 0.09 | 0.11 |
| Age | 0.11 | 0.13 | 0.10 | 0.13 | 0.11 | 0.11 | 0.12 | 0.14 |
| Citizenship status | 0.11 | 0.13 | 0.09 | 0.13 | 0.11 | 0.12 | 0.11 | 0.10 |
| Place of employment | 0.12 | 0.13 | 0.11 | 0.11 | 0.12 | 0.11 | 0.12 | 0.13 |
| Research topics | 0.19 | 0.22 | 0.17 | 0.17 | 0.20 | 0.18 | 0.20 | 0.17 |

Reported in each cell is the share of respondents who have witnessed different types of discrimination or unfair treatment by anyone in the field of economics in their current job.

Table 9B: Witnessed discrimination in the current job - by type of discrimination.

|  | No academic capital | Perceived as ethnic minority | Doctoral degree from 2010 onwards | Not in high SES |
|---|---|---|---|---|
| Have you ever witnessed discrimination or unfair treatment based on: | | | | |
| Racial/ethnic identity | 0.11 | 0.15 | 0.11 | 0.12 |
| Sex | 0.18 | 0.27 | 0.19 | 0.19 |
| Gender identity | 0.05 | 0.13 | 0.07 | 0.10 |
| Sexual orientation | 0.05 | 0.08 | 0.04 | 0.05 |
| Disability status | 0.03 | 0.04 | 0.03 | 0.03 |
| Marital status/caregiving responsibilities | 0.14 | 0.16 | 0.14 | 0.14 |
| Religion | 0.04 | 0.09 | 0.04 | 0.07 |
| Political views | 0.11 | 0.11 | 0.06 | 0.12 |
| Age | 0.10 | 0.15 | 0.09 | 0.12 |
| Citizenship status | 0.09 | 0.13 | 0.11 | 0.13 |
| Place of employment | 0.10 | 0.13 | 0.11 | 0.12 |
| Research topics | 0.17 | 0.15 | 0.18 | 0.18 |

Reported in each cell is the share of respondents who have witnessed different types of discrimination or unfair treatment by anyone in the field of economics in their current job.

Table 10A: Avoided disclosing information.

| | All | Female | Male | LGBTQ+ | Not LGBTQ+ | With disability | No disability | Ethnic minority |
|---|---|---|---|---|---|---|---|---|
| Have you ever avoided disclosing or discussing any of the following: | | | | | | | | |
| Racial/ethnic identity | 0.10 | 0.10 | 0.09 | 0.11 | 0.09 | 0.13 | 0.09 | 0.20 |
| Sex | 0.15 | 0.19 | 0.10 | 0.17 | 0.14 | 0.22 | 0.13 | 0.15 |
| Gender identity | 0.08 | 0.09 | 0.06 | 0.14 | 0.07 | 0.14 | 0.06 | 0.10 |
| Sexual orientation | 0.13 | 0.12 | 0.13 | 0.50 | 0.05 | 0.21 | 0.10 | 0.13 |
| Disability status | 0.06 | 0.07 | 0.05 | 0.09 | 0.05 | 0.20 | 0.02 | 0.06 |
| Marital status/caregiving responsibilities | 0.21 | 0.32 | 0.10 | 0.22 | 0.21 | 0.23 | 0.20 | 0.18 |
| Religion | 0.10 | 0.08 | 0.11 | 0.09 | 0.10 | 0.14 | 0.08 | 0.15 |
| Political views | 0.39 | 0.38 | 0.38 | 0.48 | 0.37 | 0.52 | 0.34 | 0.41 |
| Age | 0.13 | 0.20 | 0.07 | 0.16 | 0.13 | 0.16 | 0.12 | 0.16 |
| Citizenship status | 0.07 | 0.08 | 0.07 | 0.09 | 0.07 | 0.11 | 0.06 | 0.16 |
| Place of employment | 0.14 | 0.16 | 0.12 | 0.15 | 0.14 | 0.20 | 0.12 | 0.15 |
| Research topics | 0.21 | 0.24 | 0.16 | 0.33 | 0.18 | 0.30 | 0.18 | 0.23 |

Reported in each cell is the share of respondents who report having avoided disclosing or discussing various personal attributes to anyone in the field due to fear of negative consequences, harassment, or discrimination.

Table 10B: Avoided disclosing information.

| | No academic capital | Perceived as ethnic minority | Doctoral degree from 2010 onwards | Not in high SES |
|---|---|---|---|---|
| Have you ever avoided disclosing or discussing any of the following: | | | | |
| Racial/ethnic identity | 0.09 | 0.23 | 0.10 | 0.12 |
| Sex | 0.15 | 0.16 | 0.14 | 0.18 |
| Gender identity | 0.07 | 0.09 | 0.06 | 0.09 |
| Sexual orientation | 0.15 | 0.17 | 0.10 | 0.14 |
| Disability status | 0.08 | 0.07 | 0.06 | 0.08 |
| Marital status/caregiving responsibilities | 0.19 | 0.19 | 0.22 | 0.21 |
| Religion | 0.12 | 0.18 | 0.11 | 0.13 |
| Political views | 0.38 | 0.43 | 0.40 | 0.43 |
| Age | 0.13 | 0.17 | 0.13 | 0.17 |
| Citizenship status | 0.08 | 0.19 | 0.08 | 0.11 |
| Place of employment | 0.13 | 0.15 | 0.16 | 0.17 |
| Research topics | 0.22 | 0.24 | 0.21 | 0.25 |

Reported in each cell is the share of respondents who report having avoided disclosing or discussing various personal attributes to anyone in the field due to fear of negative consequences, harassment, or discrimination.

Table 11A: Experiences of discrimination and unfair treatment while student.

| | All | Female | Male | LGBTQ+ | Not LGBTQ+ | With disability | No disability | Ethnic minority |
|---|---|---|---|---|---|---|---|---|
| During your time as a student studying economics, have you personally experienced discrimination or unfair treatment concerning: | | | | | | | | |
| Access to research assistantships | 0.12 | 0.15 | 0.10 | 0.13 | 0.12 | 0.16 | 0.11 | 0.17 |
| Access to advisors | 0.12 | 0.15 | 0.09 | 0.11 | 0.12 | 0.15 | 0.11 | 0.13 |
| Access to quality advising | 0.15 | 0.21 | 0.10 | 0.18 | 0.15 | 0.20 | 0.14 | 0.18 |
| Job market | 0.19 | 0.19 | 0.19 | 0.16 | 0.19 | 0.22 | 0.18 | 0.23 |

Reported in each cell is the share of respondents who report having personally experienced these treatments during their time studying economics.

Table 11B: Experiences of discrimination and unfair treatment while student.

| | No academic capital | Perceived as ethnic minority | Doctoral degree from 2010 onwards | Not in high SES |
|---|---|---|---|---|
| During your time as a student studying economics, have you personally experienced discrimination or unfair treatment concerning: | | | | |
| Access to research assistantships | 0.11 | 0.19 | 0.13 | 0.17 |
| Access to advisors | 0.11 | 0.14 | 0.13 | 0.13 |
| Access to quality advising | 0.16 | 0.18 | 0.18 | 0.18 |
| Job market | 0.22 | 0.27 | 0.24 | 0.24 |

Reported in each cell is the share of respondents who report having personally experienced these treatments during their time studying economics.

Table 12A: Experiences of discrimination in academia.

| | All | Female | Male | LGBTQ+ | Not LGBTQ+ | With disability | No disability | Ethnic minority |
|---|---|---|---|---|---|---|---|---|
| Have you personally experienced discrimination or unfair treatment with regard to: | | | | | | | | |
| Hiring decisions | 0.24 | 0.23 | 0.25 | 0.25 | 0.24 | 0.28 | 0.22 | 0.31 |
| Promotion decisions | 0.19 | 0.24 | 0.13 | 0.15 | 0.19 | 0.21 | 0.18 | 0.23 |
| Compensation | 0.19 | 0.27 | 0.12 | 0.17 | 0.19 | 0.23 | 0.18 | 0.19 |
| Teaching assignments | 0.15 | 0.21 | 0.08 | 0.14 | 0.15 | 0.16 | 0.14 | 0.17 |
| Service obligations | 0.18 | 0.29 | 0.07 | 0.13 | 0.19 | 0.20 | 0.17 | 0.18 |
| Committee or decision-making body membership | 0.17 | 0.24 | 0.10 | 0.14 | 0.17 | 0.20 | 0.16 | 0.17 |
| Access to time and funding to attend conferences and seminars | 0.12 | 0.14 | 0.10 | 0.09 | 0.13 | 0.14 | 0.12 | 0.16 |
| Access to graduate student researchers or research assistants | 0.11 | 0.14 | 0.08 | 0.09 | 0.11 | 0.13 | 0.10 | 0.15 |
| Course evaluations | 0.16 | 0.28 | 0.05 | 0.17 | 0.15 | 0.20 | 0.14 | 0.14 |
| Publishing decisions | 0.20 | 0.22 | 0.18 | 0.18 | 0.20 | 0.23 | 0.19 | 0.23 |
| Funding decisions | 0.14 | 0.15 | 0.13 | 0.12 | 0.14 | 0.18 | 0.13 | 0.17 |
| Sabbatical time | 0.04 | 0.05 | 0.03 | 0.02 | 0.04 | 0.04 | 0.04 | 0.05 |
| Access to potential coauthors | 0.11 | 0.18 | 0.05 | 0.11 | 0.11 | 0.17 | 0.09 | 0.16 |
| Invitations to participate in research conferences, associations and networks | 0.18 | 0.18 | 0.18 | 0.18 | 0.19 | 0.21 | 0.17 | 0.24 |
| Seminar invitations | 0.15 | 0.15 | 0.14 | 0.10 | 0.15 | 0.16 | 0.14 | 0.21 |
| Hostility during presentations/talks | 0.23 | 0.32 | 0.16 | 0.19 | 0.24 | 0.28 | 0.22 | 0.25 |

Reported in each cell is the share of respondents who report having personally experienced discrimination or unfair treatment in these aspects of work over the last ten years in the field of economics in Europe.

Table 12B: Experiences of discrimination in academia.

| | No academic capital | Perceived as ethnic minority | Doctoral degree from 2010 onwards | Not in high SES |
|---|---|---|---|---|
| Have you personally experienced discrimination or unfair treatment with regard to: | | | | |
| Hiring decisions | 0.26 | 0.37 | 0.30 | 0.29 |
| Promotion decisions | 0.21 | 0.22 | 0.21 | 0.21 |
| Compensation | 0.17 | 0.22 | 0.21 | 0.20 |
| Teaching assignments | 0.15 | 0.20 | 0.16 | 0.18 |
| Service obligations | 0.17 | 0.20 | 0.18 | 0.17 |
| Committee or decision-making body membership | 0.16 | 0.22 | 0.17 | 0.18 |
| Access to time and funding to attend conferences and seminars | 0.12 | 0.20 | 0.15 | 0.14 |
| Access to graduate student researchers or research assistants | 0.11 | 0.19 | 0.10 | 0.14 |
| Course evaluations | 0.13 | 0.18 | 0.19 | 0.18 |
| Publishing decisions | 0.23 | 0.26 | 0.25 | 0.25 |
| Funding decisions | 0.17 | 0.20 | 0.16 | 0.18 |
| Sabbatical time | 0.04 | 0.06 | 0.02 | 0.04 |
| Access to potential coauthors | 0.12 | 0.18 | 0.15 | 0.13 |
| Invitations to participate in research conferences, associations and networks | 0.17 | 0.28 | 0.24 | 0.21 |
| Seminar invitations | 0.14 | 0.26 | 0.19 | 0.17 |
| Hostility during presentations/talks | 0.24 | 0.24 | 0.28 | 0.26 |

Reported in each cell is the share of respondents who report having personally experienced discrimination or unfair treatment in these aspects of work over the last ten years in the field of economics in Europe.

Table 13A: Actions taken to avoid possible harassment, discrimination, or unfair treatment.

| | All | Female | Male | LGBTQ+ | Not LGBTQ+ | With disability | No disability | Ethnic minority |
|---|---|---|---|---|---|---|---|---|
| **Have you ever done any of the following to avoid possible harassment, discrimination, or unfair or disrespectful treatment:** | | | | | | | | |
| Not applied for or accepted admission at a particular grad school | 0.09 | 0.11 | 0.06 | 0.15 | 0.08 | 0.14 | 0.07 | 0.11 |
| Paused or ceased enrollment at a particular grad school | 0.03 | 0.04 | 0.02 | 0.03 | 0.03 | 0.03 | 0.03 | 0.04 |
| Not applied for or taken a particular employment position | 0.18 | 0.22 | 0.12 | 0.21 | 0.17 | 0.20 | 0.17 | 0.19 |
| Not applied for or taken a promotion at your place of employment | 0.08 | 0.10 | 0.06 | 0.07 | 0.08 | 0.08 | 0.08 | 0.13 |
| Left a particular employment position | 0.13 | 0.17 | 0.09 | 0.15 | 0.13 | 0.19 | 0.12 | 0.17 |
| Not presented your question, idea, or view at your school or place of work | 0.40 | 0.54 | 0.26 | 0.40 | 0.40 | 0.48 | 0.37 | 0.43 |
| Not participated in a conference | 0.20 | 0.24 | 0.16 | 0.19 | 0.20 | 0.24 | 0.18 | 0.23 |
| Not spoken at a conference or during a seminar presentation | 0.42 | 0.56 | 0.30 | 0.39 | 0.43 | 0.55 | 0.38 | 0.43 |
| Not made a professional visit to a particular place | 0.13 | 0.17 | 0.09 | 0.13 | 0.13 | 0.22 | 0.09 | 0.13 |
| Not attended social events after class, at work, or at conferences | 0.30 | 0.43 | 0.17 | 0.32 | 0.29 | 0.41 | 0.26 | 0.35 |
| Changed the topic, content, or method of a class you teach | 0.10 | 0.11 | 0.09 | 0.10 | 0.10 | 0.13 | 0.09 | 0.13 |
| Changed the content, method, or conclusions of a research paper | 0.14 | 0.15 | 0.12 | 0.17 | 0.13 | 0.18 | 0.13 | 0.19 |
| Not started or continued research in a particular field | 0.23 | 0.28 | 0.18 | 0.28 | 0.22 | 0.33 | 0.20 | 0.34 |

Reported in each cell is the share of respondents who report having taken the listed action to avoid possible harassment, discrimination, or unfair or disrespectful treatment over the last ten years in the field of economics.

Table 13B: Actions taken to avoid possible harassment, discrimination, or unfair treatment.

| | No academic capital | Perceived as ethnic minority | Doctoral degree from 2010 onwards | Not in high SES |
|---|---|---|---|---|
| Have you ever done any of the following to avoid possible harassment, discrimination, or unfair or disrespectful treatment: | | | | |
| Not applied for or accepted admission at a particular grad school | 0.08 | 0.13 | 0.10 | 0.11 |
| Paused or ceased enrollment at a particular grad school | 0.01 | 0.03 | 0.03 | 0.03 |
| Not applied for or taken a particular employment position | 0.19 | 0.19 | 0.26 | 0.18 |
| Not applied for or taken a promotion at your place of employment | 0.09 | 0.12 | 0.09 | 0.12 |
| Left a particular employment position | 0.12 | 0.18 | 0.15 | 0.14 |
| Not presented your question, idea, or view at your school or place of work | 0.41 | 0.44 | 0.37 | 0.46 |
| Not participated in a conference | 0.20 | 0.28 | 0.21 | 0.23 |
| Not spoken at a conference or during a seminar presentation | 0.40 | 0.43 | 0.43 | 0.43 |
| Not made a professional visit to a particular place | 0.13 | 0.16 | 0.14 | 0.14 |
| Not attended social events after class, at work, or at conferences | 0.30 | 0.38 | 0.28 | 0.32 |
| Changed the topic, content, or method of a class you teach | 0.13 | 0.16 | 0.10 | 0.12 |
| Changed the content, method, or conclusions of a research paper | 0.17 | 0.22 | 0.15 | 0.18 |
| Not started or continued research in a particular field | 0.23 | 0.37 | 0.26 | 0.26 |

Reported in each cell is the share of respondents who report having taken the listed action to avoid possible harassment, discrimination, or unfair or disrespectful treatment over the last ten years in the field of economics.

Table 14A: Experiences of Exclusion and Harassment.

| | All | Female | Male | LGBTQ+ | Not LGBTQ+ | With disability | No disability | Ethnic minority |
|---|---|---|---|---|---|---|---|---|
| Have you ever experienced any of the following: | | | | | | | | |
| Felt socially excluded at a meeting or event in the field | 0.59 | 0.69 | 0.50 | 0.61 | 0.58 | 0.68 | 0.56 | 0.61 |
| Felt disrespected by your economist colleagues | 0.56 | 0.69 | 0.44 | 0.54 | 0.57 | 0.62 | 0.54 | 0.59 |
| Felt that your work was not taken as seriously as that of your economist colleagues | 0.65 | 0.75 | 0.57 | 0.60 | 0.66 | 0.68 | 0.64 | 0.59 |
| Felt that the subject or methodology of your research was not taken as seriously as that of your economist colleagues | 0.56 | 0.64 | 0.50 | 0.52 | 0.57 | 0.62 | 0.54 | 0.53 |
| Another economist or economics student displayed inappropriate behavior towards you, including offensive sexual remarks, remarks about your appearance OR body, inappropriate gestures | 0.26 | 0.42 | 0.10 | 0.26 | 0.25 | 0.36 | 0.22 | 0.27 |
| Another economist or economics student made unwanted attempts to establish a dating, romantic, or sexual relationship with you despite your efforts to discourage it | 0.11 | 0.20 | 0.03 | 0.11 | 0.11 | 0.13 | 0.10 | 0.10 |
| Another economist or economics student threatened you with retaliation for not being romantically or sexually cooperative, or implied you'd be treated better if you were sexually cooperative. | 0.02 | 0.04 | 0.00 | 0.01 | 0.02 | 0.03 | 0.02 | 0.02 |
| Another economist or economics student stalked you | 0.06 | 0.07 | 0.05 | 0.05 | 0.06 | 0.05 | 0.06 | 0.07 |
| Another economist or economics student attempted to sexually assault you | 0.02 | 0.03 | 0.00 | 0.02 | 0.02 | 0.03 | 0.01 | 0.03 |
| Another economist or economics student sexually assaulted you | 0.01 | 0.03 | 0.00 | 0.03 | 0.01 | 0.02 | 0.01 | 0.03 |

Reported in each cell is the share of respondents that report having personally experienced the stated treatment over the last ten years in the field of economics in Europe.

Table 14B: Experiences of Exclusion and Harassment.

| | No academic capital | Perceived as ethnic minority | Doctoral degree from 2010 onwards | Not in high SES |
|---|---|---|---|---|
| Have you ever experienced any of the following: | | | | |
| Felt socially excluded at a meeting or event in the field | 0.60 | 0.61 | 0.63 | 0.66 |
| Felt disrespected by your economist colleagues | 0.54 | 0.58 | 0.58 | 0.59 |
| Felt that your work was not taken as seriously as that of your economist colleagues | 0.65 | 0.61 | 0.69 | 0.68 |
| Felt that the subject or methodology of your research was not taken as seriously as that of your economist colleagues | 0.55 | 0.55 | 0.59 | 0.62 |
| Another economist or economics student displayed inappropriate behavior towards you, including offensive sexual remarks, remarks about your appearance OR body, inappropriate gestures | 0.24 | 0.27 | 0.27 | 0.27 |
| Another economist or economics student made unwanted attempts to establish a dating, romantic, or sexual relationship with you despite your efforts to discourage it | 0.11 | 0.14 | 0.12 | 0.12 |
| Another economist or economics student threatened you with retaliation for not being romantically or sexually cooperative, or implied you'd be treated better if you were sexually cooperative. | 0.01 | 0.02 | 0.02 | 0.02 |
| Another economist or economics student stalked you | 0.06 | 0.08 | 0.06 | 0.06 |
| Another economist or economics student attempted to sexually assault you | 0.02 | 0.04 | 0.03 | 0.03 |
| Another economist or economics student sexually assaulted you | 0.02 | 0.03 | 0.02 | 0.02 |

Reported in each cell is the share of respondents who report having personally experienced the stated treatment over the last ten years in the field of economics in Europe.

Table 15: Experiences of sexual assault.

| | Perc. | N |
|---|---|---|
| **Where?** | | |
| At my university | 0.32 | 6 |
| At another university | 0.05 | 1 |
| At my non-academic workplace | 0.05 | 1 |
| At a professional conference or meeting | 0.26 | 5 |
| During a conference or meeting, but not at the conference or meeting itself | 0.26 | 5 |
| Online (via email or other electronic media) | 0.00 | 0 |
| Somewhere else | 0.00 | 0 |
| **Who?** | | |
| My professor/boss/someone with authority | 0.53 | 10 |
| A co-worker at my institution or place of employment | 0.11 | 2 |
| Another economist or student that I know | 0.26 | 5 |
| Another economist or student that I do not know | 0.05 | 1 |
| Someone else | 0.00 | 0 |
| Do not know the identity or status of this person | 0.00 | 0 |
| **Told anyone?** | | |
| Yes | 0.37 | 7 |
| No | 0.63 | 12 |
| **Who did you tell first?** | | |
| A colleague | 0.57 | 4 |
| A relevant administration official at the workplace/university/institution | 0.14 | 1 |
| The police | 0.00 | 0 |
| Someone else | 0.29 | 2 |
| **If you did not make a report to some authority, why not?** | | N=19 |
| Didn't know who the right person was | 0.21 | 4 |
| Were concerned the situation would not be kept confidential | 0.37 | 7 |
| Did not need/want any assistance or any action taken | 0.16 | 3 |
| Were concerned the process and/or the outcome would be too difficult | 0.21 | 4 |
| Were concerned about retribution from the person who did this and/or others over reporting | 0.32 | 6 |
| Did not think that the person who did this to you will be held accountable | 0.53 | 10 |
| Thought the consequences for that person would be too harsh relative to the extent of the misconduct | 0.16 | 3 |
| **Did the experience lead you to…** | | N=19 |
| File an official charge of complaint | 0.11 | 2 |
| File an official charge with police | 0.00 | 0 |
| Consider leaving a project, committee, program | 0.32 | 6 |
| Be less productive or effective in your work | 0.53 | 10 |
| Consider leaving your position | 0.26 | 5 |
| Take leave, sick time, miss work unexpectedly, or other similar time away from work | 0.11 | 2 |
| Consider not attending future conferences | 0.32 | 6 |
| Consider leaving the field of economic research | 0.32 | 6 |
| Consider leaving academia entirely | 0.32 | 6 |
| Consider taking legal action | 0.16 | 3 |

Table 16: General climate - regression analysis.

| | Q1 | Q2 | Q3 | Q4 | Q5 | Q6 | Q7 | Q8 | Q9 | Q10 | Q11 | Q12 | Q13 | Q14 | Q15 | Q16 |
|---|---|---|---|---|---|---|---|---|---|---|---|---|---|---|---|---|
| Female | -0.118* | -0.079 | 0.003 | -0.163*** | -0.027 | -0.124** | -0.016 | -0.074 | 0.105 | 0.179*** | 0.047 | 0.033 | 0.001 | -0.023 | -0.112* | 0.049 |
| | [0.065] | [0.058] | [0.064] | [0.056] | [0.063] | [0.053] | [0.062] | [0.053] | [0.064] | [0.055] | [0.058] | [0.052] | [0.044] | [0.06] | [0.065] | [0.059] |
| LGBTQ+ | -0.036 | -0.048 | -0.047 | -0.01 | -0.044 | -0.047 | -0.001 | 0.002 | 0.062 | -0.022 | -0.004 | -0.049 | -0.014 | -0.028 | 0.03 | -0.031 |
| | [0.048] | [0.043] | [0.047] | [0.042] | [0.047] | [0.039] | [0.046] | [0.04] | [0.048] | [0.041] | [0.043] | [0.039] | [0.032] | [0.045] | [0.048] | [0.044] |
| Disability | -0.018 | -0.045 | -0.026 | -0.051 | -0.098** | -0.088** | -0.106*** | -0.079** | 0.12*** | 0.111*** | -0.059 | -0.067* | -0.023 | -0.065* | 0.111*** | 0.069* |
| | [0.042] | [0.038] | [0.041] | [0.037] | [0.041] | [0.035] | [0.04] | [0.035] | [0.042] | [0.036] | [0.038] | [0.034] | [0.029] | [0.039] | [0.042] | [0.038] |
| AgeOver39 | 0.133** | 0.064 | 0.104* | 0.065 | 0.077 | 0.081* | 0.08 | 0.064 | -0.121** | -0.069 | 0.079 | 0.032 | 0.08** | 0.114** | -0.058 | -0.043 |
| | [0.056] | [0.05] | [0.055] | [0.048] | [0.054] | [0.046] | [0.053] | [0.046] | [0.055] | [0.048] | [0.05] | [0.045] | [0.038] | [0.052] | [0.056] | [0.051] |
| AgeUnder39 | 0.107* | 0.104** | -0.01 | 0.057 | 0.046 | 0.088* | 0.055 | 0.053 | -0.079 | -0.104** | 0.039 | 0.042 | -0.082** | -0.106** | -0.007 | -0.087* |
| | [0.055] | [0.049] | [0.054] | [0.048] | [0.054] | [0.045] | [0.053] | [0.046] | [0.055] | [0.047] | [0.05] | [0.045] | [0.037] | [0.051] | [0.056] | [0.05] |
| Political Lean | -0.041 | 0.003 | -0.02 | -0.021 | -0.023 | 0.004 | -0.025 | 0.024 | -0.035 | 0.021 | -0.034 | -0.01 | -0.062** | -0.089** | 0.009 | 0.029 |
| | [0.041] | [0.037] | [0.04] | [0.036] | [0.04] | [0.034] | [0.039] | [0.034] | [0.041] | [0.035] | [0.037] | [0.033] | [0.028] | [0.038] | [0.041] | [0.037] |
| Married | 0.07 | 0.039 | 0.053 | -0.012 | 0.106* | 0.04 | 0.033 | 0.025 | -0.024 | -0.027 | 0.053 | 0.041 | 0.088** | 0.097* | -0.046 | -0.058 |
| | [0.058] | [0.052] | [0.057] | [0.05] | [0.057] | [0.048] | [0.055] | [0.048] | [0.058] | [0.05] | [0.052] | [0.047] | [0.039] | [0.054] | [0.058] | [0.053] |
| Has Dependents | 0.045 | 0.037 | 0.086* | 0.015 | 0.071 | -0.003 | 0.132*** | 0.041 | 0.06 | 0.034 | 0.11** | 0.057 | 0.023 | 0.054 | 0.033 | -0.008 |
| | [0.052] | [0.046] | [0.051] | [0.045] | [0.05] | [0.043] | [0.049] | [0.043] | [0.051] | [0.044] | [0.047] | [0.042] | [0.035] | [0.048] | [0.052] | [0.047] |
| Female & Married | -0.028 | -0.016 | -0.078 | 0.126* | -0.118 | 0.073 | -0.042 | 0.053 | 0.037 | -0.066 | -0.015 | -0.007 | -0.049 | -0.066 | 0.069 | -0.047 |
| | [0.08] | [0.071] | [0.078] | [0.069] | [0.078] | [0.065] | [0.076] | [0.066] | [0.079] | [0.068] | [0.072] | [0.064] | [0.054] | [0.074] | [0.08] | [0.072] |
| Female & Has Dependents | -0.047 | -0.103 | -0.044 | -0.051 | -0.079 | -0.067 | -0.141** | -0.083 | 0.024 | 0.062 | -0.11* | -0.133** | -0.032 | -0.056 | 0.06 | 0.084 |
| | [0.072] | [0.064] | [0.07] | [0.062] | [0.07] | [0.059] | [0.068] | [0.059] | [0.071] | [0.061] | [0.065] | [0.058] | [0.048] | [0.067] | [0.072] | [0.065] |
| Ethnic Group | -0.07* | -0.024 | -0.085** | -0.083** | -0.122*** | -0.043 | -0.112*** | -0.048 | 0.105** | -0.004 | -0.018 | -0.029 | -0.048* | -0.021 | 0.015 | 0.044 |
| | [0.041] | [0.037] | [0.041] | [0.036] | [0.04] | [0.034] | [0.04] | [0.034] | [0.041] | [0.035] | [0.037] | [0.034] | [0.028] | [0.038] | [0.042] | [0.038] |
| Socioeconomic Status | 0.035 | 0.108*** | 0.079** | 0.096*** | 0.099*** | 0.096*** | 0.061* | 0.079*** | -0.038 | -0.095*** | 0.042 | 0.059** | 0.036 | 0.052 | -0.121*** | -0.14*** |
| | [0.036] | [0.032] | [0.035] | [0.031] | [0.035] | [0.029] | [0.034] | [0.03] | [0.035] | [0.031] | [0.032] | [0.029] | [0.024] | [0.033] | [0.036] | [0.033] |
| (Constant) | 0.506 | 0.627 | 0.535 | 0.738 | 0.523 | 0.704 | 0.593 | 0.695 | 0.41 | 0.274 | 0.617 | 0.728 | 0.129 | 0.343 | 0.508 | 0.407 |
| | [0.076] | [0.068] | [0.075] | [0.066] | [0.074] | [0.063] | [0.073] | [0.063] | [0.076] | [0.065] | [0.069] | [0.062] | [0.051] | [0.071] | [0.077] | [0.069] |
| R Squared | .055 | .055 | .054 | .049 | .080 | .065 | .061 | .040 | .059 | .073 | .030 | .031 | .098 | .106 | .034 | .049 |

All variables are 0/1. Standard errors in brackets; *** p<0.01, ** p<0.05, * p<0.1.

Q1. I am satisfied with the overall professional climate within the field of economics.

Q2. I am satisfied with the overall professional climate at my institution/place of employment.

Q3. I feel valued within the field of economics.

Q4. I feel valued at my institution/place of employment.

Q5. I always feel included socially within the field of economics.

Q6. I always feel included socially at my institution/place of employment.

Q7. I always feel included intellectually within the field of economics.

Q8. I always feel included intellectually at my institution/place of employment.

Q9. I feel I have been discriminated against within the field of economics.

Q10. I feel I have been discriminated against at my institution/place of employment.

Q11. The work that I do is valued within the field of economics.

Q12. The work that I do is valued at my institution/place of employment.

Q13. I think I have a great deal of power within the field of economics.

Q14. I think I have a great deal of power at my institution/place of employment.

Q15. My ideas and opinions are often ignored within the field of economics.

Table 17: Experiences of exclusion and harassment - regression analysis.

| | Socially excluded | Disrespected | Work not taken seriously | Subject not taken seriously | Inappropriate behaviour | Harassed |
|---|---|---|---|---|---|---|
| Female | 0.203*** | 0.136** | 0.044 | 0.022 | 0.23*** | 0.168*** |
| | [0.064] | [0.064] | [0.062] | [0.065] | [0.054] | [0.047] |
| LGBTQ+ | -0.016 | -0.025 | -0.055 | -0.062 | 0.001 | 0.004 |
| | [0.047] | [0.048] | [0.046] | [0.048] | [0.04] | [0.035] |
| Disability | 0.113*** | 0.062 | 0.046 | 0.049 | 0.105*** | 0.004 |
| | [0.042] | [0.042] | [0.041] | [0.042] | [0.035] | [0.031] |
| AgeOver39 | -0.086 | 0.047 | -0.001 | -0.007 | -0.033 | 0.027 |
| | [0.055] | [0.055] | [0.053] | [0.056] | [0.047] | [0.04] |
| AgeUnder39 | 0.005 | 0.013 | 0.02 | -0.004 | 0.02 | 0.005 |
| | [0.055] | [0.055] | [0.053] | [0.056] | [0.046] | [0.04] |
| Political Lean | 0.014 | 0.05 | 0.032 | 0.059 | 0.018 | 0 |
| | [0.041] | [0.041] | [0.039] | [0.041] | [0.034] | [0.03] |
| Married | 0.018 | -0.017 | -0.059 | -0.035 | -0.005 | -0.05 |
| | [0.057] | [0.058] | [0.056] | [0.058] | [0.049] | [0.042] |
| Has Dependents | -0.014 | -0.112** | -0.049 | -0.12** | -0.035 | 0.008 |
| | [0.051] | [0.051] | [0.05] | [0.052] | [0.043] | [0.037] |
| Female & Married | -0.063 | 0.055 | 0.108 | 0.027 | 0.041 | 0.004 |
| | [0.079] | [0.079] | [0.076] | [0.08] | [0.067] | [0.057] |
| Female & Has Dependents | 0.045 | 0.13* | 0.139** | 0.208*** | 0.099* | -0.021 |
| | [0.071] | [0.071] | [0.069] | [0.072] | [0.06] | [0.052] |
| Ethnic Group | 0.002 | 0.042 | -0.07* | -0.047 | 0.017 | -0.001 |
| | [0.041] | [0.041] | [0.04] | [0.042] | [0.035] | [0.03] |
| Socioeconomic Status | -0.128*** | -0.065* | -0.089** | -0.125*** | -0.047 | -0.039 |
| | [0.035] | [0.036] | [0.034] | [0.036] | [0.03] | [0.026] |
| (Constant) | 0.583 | 0.481 | 0.674 | 0.637 | 0.133 | 0.119 |
| | [0.075] | [0.076] | [0.073] | [0.077] | [0.064] | [0.055] |
| R Squared | .068 | .076 | .063 | .052 | .148 | .057 |

All variables are 0/1. Standard errors in brackets; *** $p<0.01$, ** $p<0.05$, * $p<0.1$.

## 3. Differences across countries

Figure 2 presents summary statistics of the socio-demographic characteristics, revealing significant differences across countries. One of the most interesting findings is the distribution of LGBTQ+ respondents. This share ranges from 6% in Nordic countries to 26.4% in the Netherlands and Belgium. Notably, it is much higher among respondents who preferred not to disclose their country of employment, at 45.6%. One possible explanation is that individuals who answered "prefer not to answer" to the question about their sexual orientation are identified as LGBTQ+. These represent 41.6% (62 out of 149) of the LGBTQ+ individuals.

The percentage of female respondents varies from 43.5% in the UK and Ireland to 53.3% in Eastern Europe. France has the highest rate of respondents who do not belong to a high socio-economic status, at approximately 78%. In contrast, Eastern Europe has the lowest, with over 53% of respondents in a high socio-economic status.

Regarding respondents with a disability, Eastern Europe has the highest share, at around 40%, while other regions have lower percentages, with Italy having the lowest share at 12.7%. In terms of age, the countries with the highest proportion of respondents under 39 years of age are France (70.7%), the UK/Ireland (67%), and Germany (60%). Conversely, this proportion is lowest in Eastern Europe at 26.7%.

Eastern Europe also has the lowest percentage of respondents who belong to an ethnic minority group, at around 6.7%, whereas this share is significantly higher in Iberia, at approximately 28.6%.

Figure 2: Socio-demographics across countries.

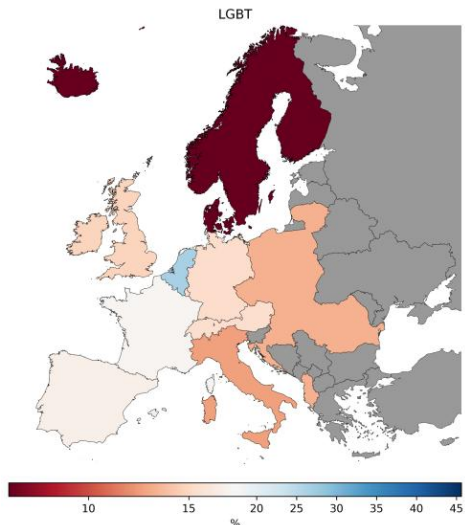
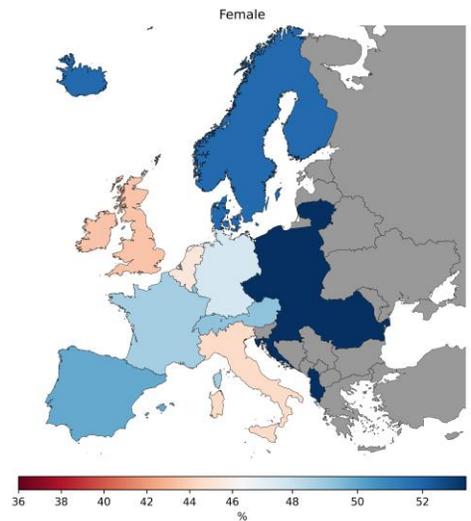
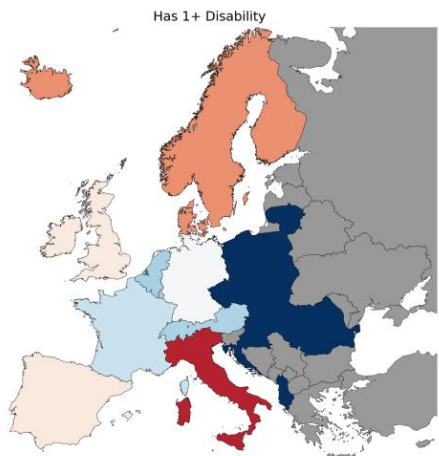
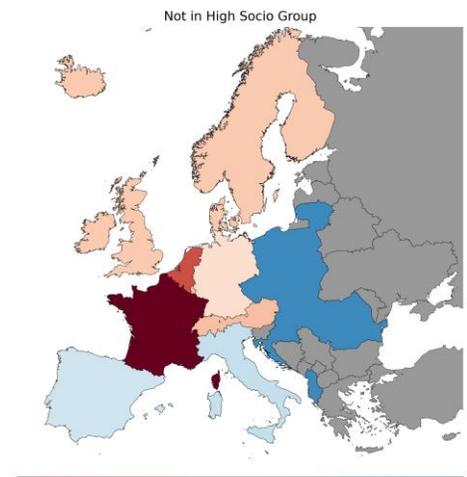
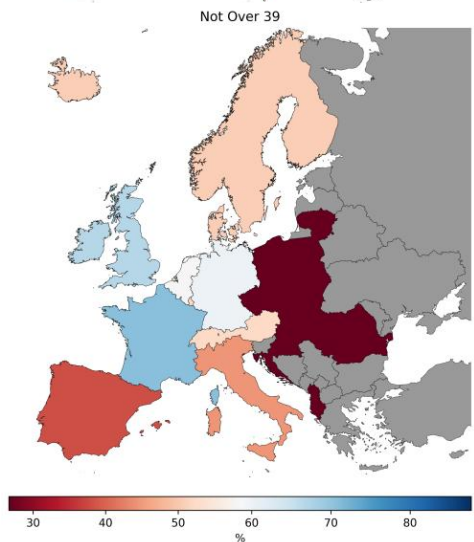
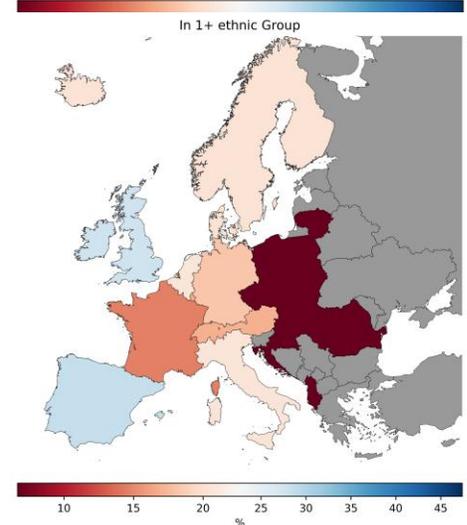

Based on the information collected in the first part of the survey, we created an indicator of the "general climate." This indicator was constructed using responses to the first 16 questions related to the climate within the economics profession. These questions included statements like "I feel valued within the field of economics" and "I always feel included socially at my institution/place of employment" which aim to assess various workplace aspects that can influence whether an individual feels valued, respected, and treated fairly. The response options for each of these 16 questions were presented on a 6-point scale, ranging from "strongly agree" to "strongly disagree". Each question's responses were ordered from most positive (6) to most negative (1), and an average score was calculated for each country. This average was weighted by the number of respondents selecting each option. Figure 3 displays the distribution of this indicator across different countries.

This shows a relatively wide spread in the overall climate score across countries. The score ranges from 3.67 in Italy to 4.18 in the Nordic countries (Denmark, Finland, Iceland, Norway, and Sweden), with an overall average score across all respondents of 3.85. Other regions with an overall climate score higher than the average are Iberia (Spain and Portugal), Germany, and German-speaking countries (Switzerland, Austria, and Luxembourg).

Figure 3: Summary of general climate across countries

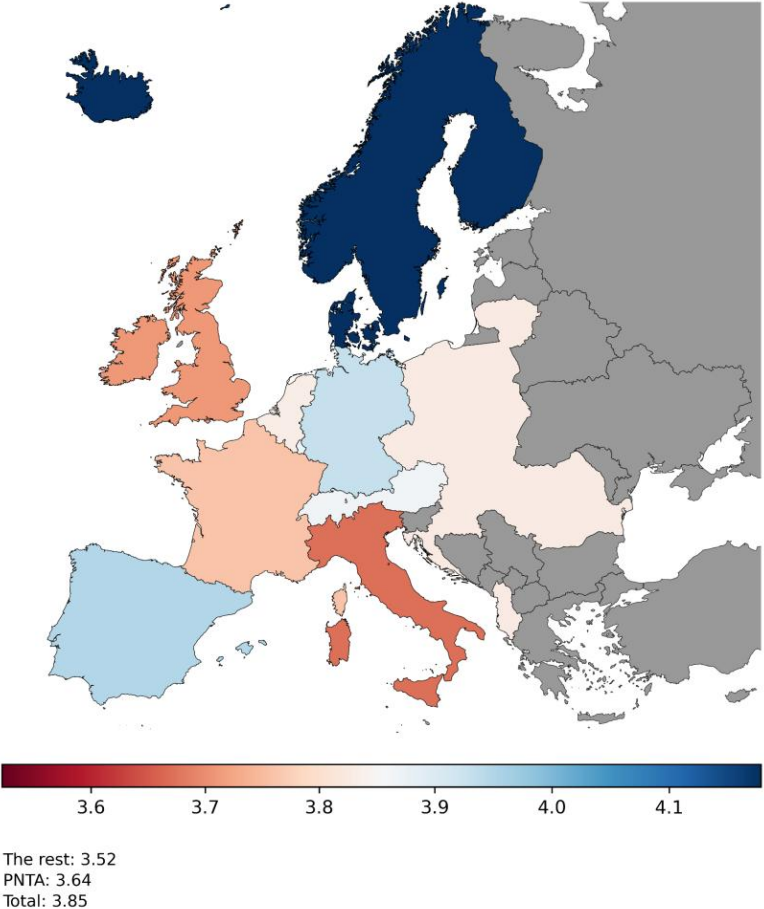

A closer examination of direct experiences of discrimination reveals notable differences across countries. The survey includes four questions designed to capture respondents' experiences of discrimination within the field of economics. Specifically, the survey asks whether, in the last 10 years in the field of economics in Europe, respondents have experienced any of the following: (1) "felt socially excluded at a meeting or event," (2) "felt disrespected by your economist colleagues," (3) "felt that your work was not taken as seriously as that of your economist colleagues," and (4) "felt that the subject or methodology of your research was not taken as seriously as that of your economist colleagues."

Figure 5 presents the average across the shares of respondents who answered "yes" to these questions. The UK and Ireland stand out as the region with the highest share of respondents who have experienced discrimination (57.4%), followed closely by Italy (56.3%) and German-speaking countries (55.9%). On the other hand, Spain and Portugal have the lowest share of respondents reporting such experiences (45.9%).

Figure 5: Experiences of discrimination.

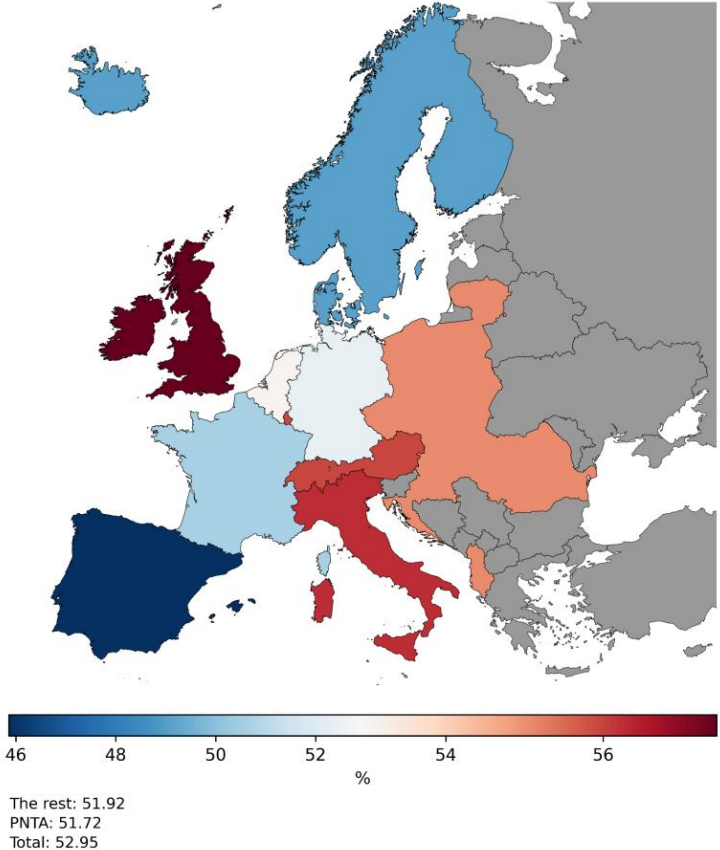

Another set of questions is designed to measure the incidence and types of sexual harassment, coercion, and assault that individuals might experience within the professional context of economics. These questions include items such as "another economist or economics student displayed inappropriate behavior towards you, including offensive sexual remarks, comments about your appearance or body, or inappropriate gestures," and "another economist or economics student threatened you with retaliation for not being romantically or sexually cooperative, or implied you'd be treated better if you were sexually cooperative."

Figure 6 presents the average share of respondents who answered "yes" to these questions. The region with the highest incidence of such experiences is Spain and Portugal (11.1%), while the lowest incidence is found in the Nordic countries (5%).

Figure 6: Experiences of sexual harassment, coercion, and assault

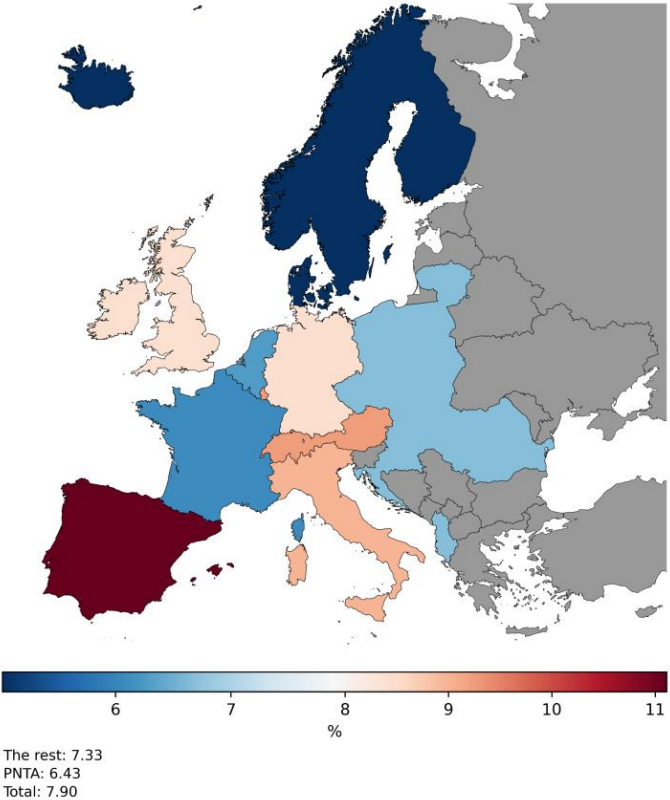

Figure A shows a breakdown of all the items in Figures 5 and 6.

Figure A: In your last 10 years in the field of economics in Europe, have you ever experienced any of the following?

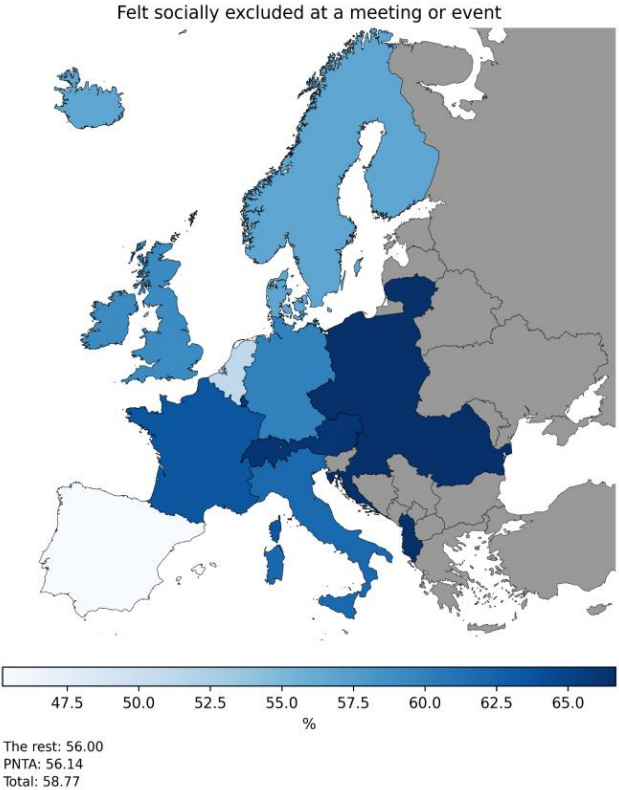

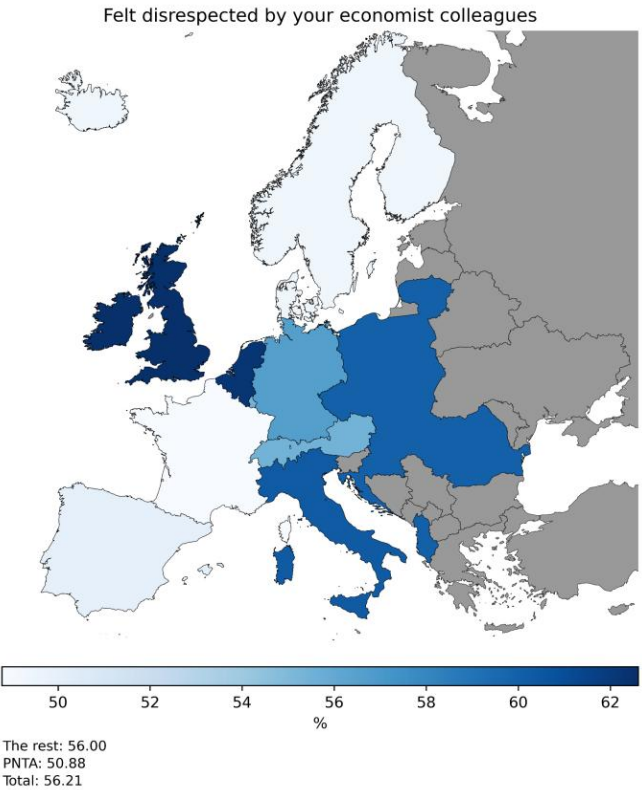

Felt that your work was not taken as seriously as that of your economist colleagues

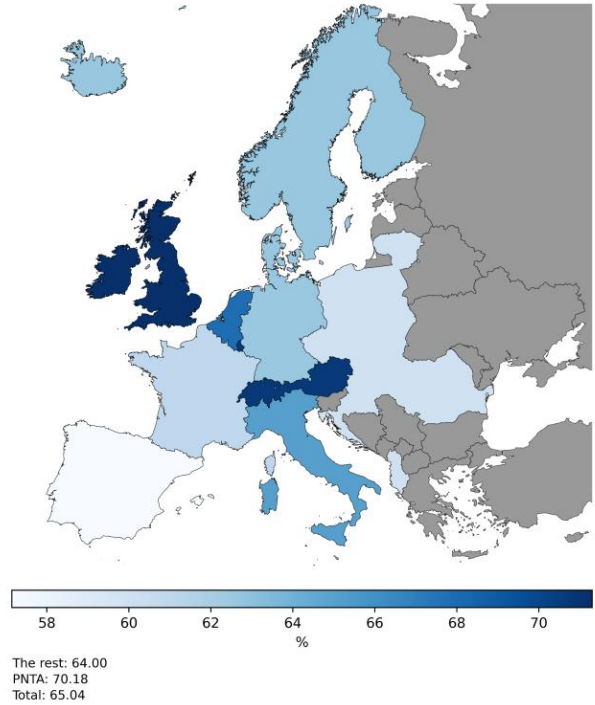

Felt that the subject or methodology of your research was not taken as seriously as that of your economist colleagues

Another economist or economics student displayed inappropriate behavior towards you, including offensive sexual remarks…

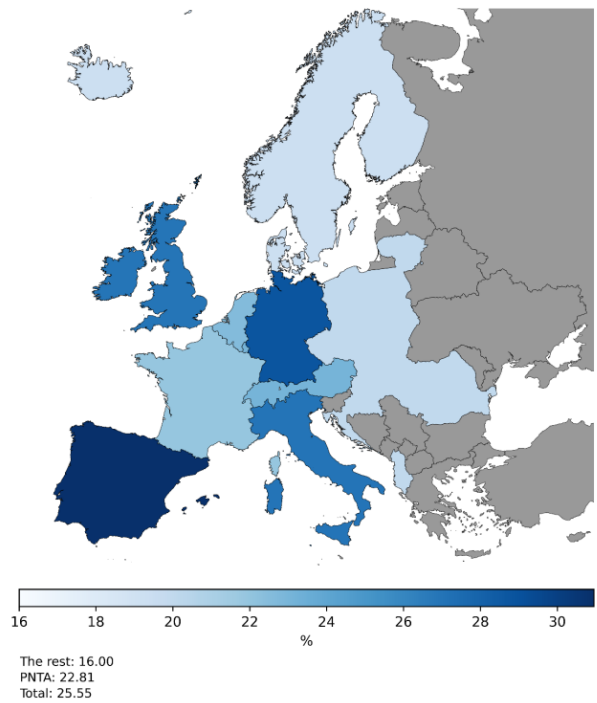

Another economist or economics student made unwanted attempts to establish a dating, romantic, or sexual relationship with you...

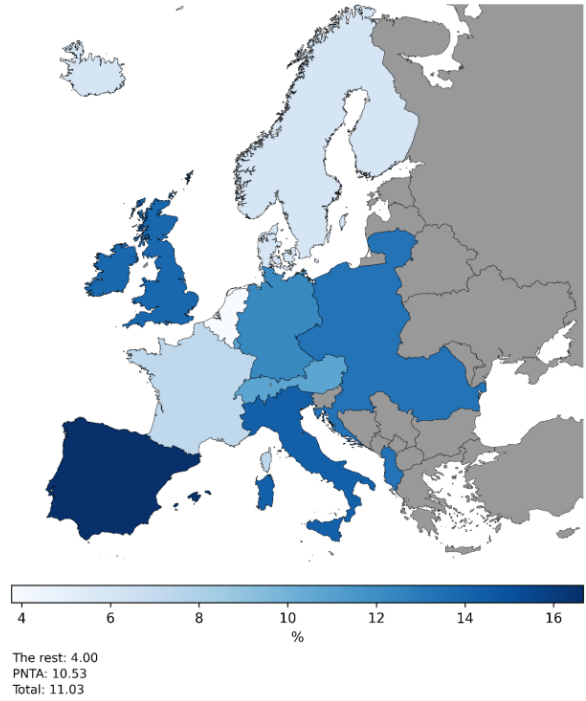

The rest: 4.00
PNTA: 10.53
Total: 11.03

Another economist or economics student threatened you with retaliation for not being romantically or sexually cooperative...

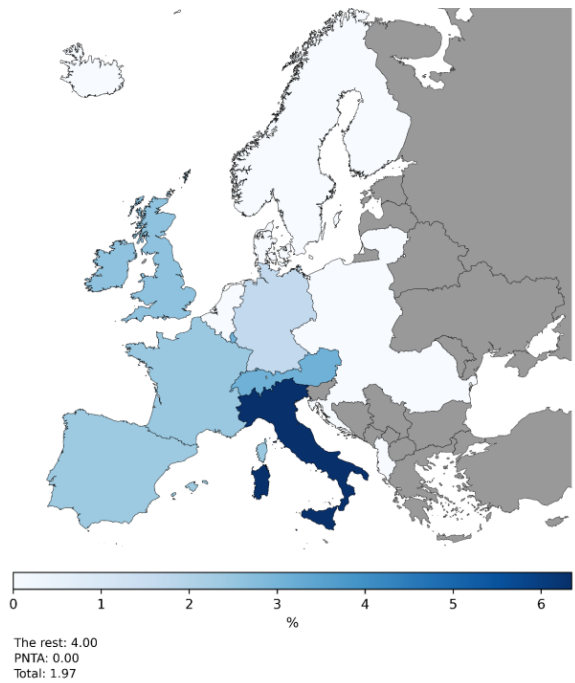

The rest: 4.00
PNTA: 0.00
Total: 1.97

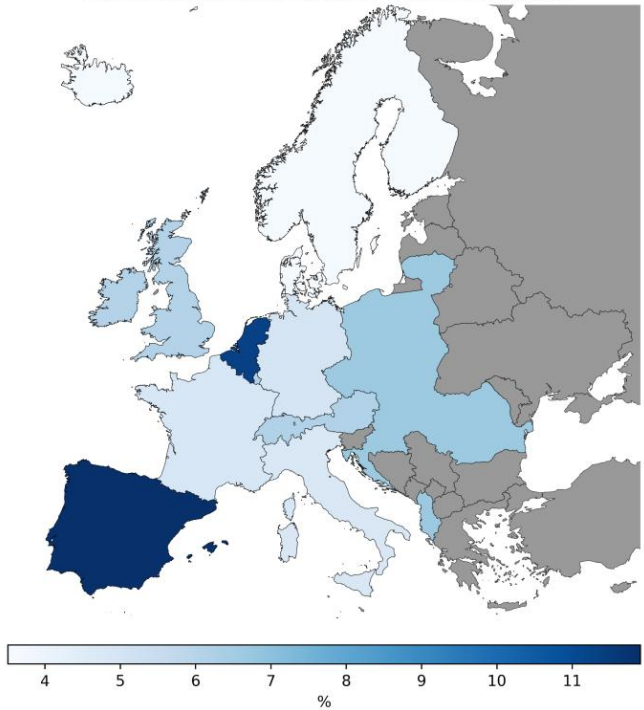

Another economist or economics student stalked you

The rest: 8.00
PNTA: 3.51
Total: 5.81

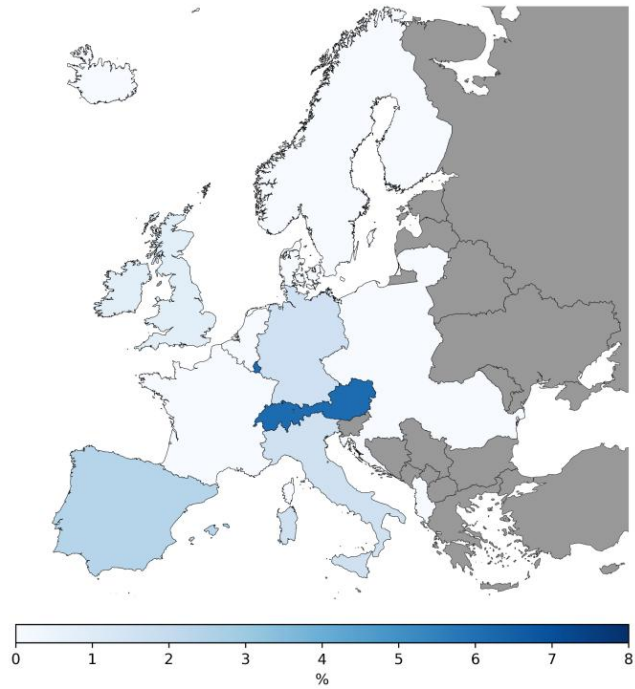

Another economist or economics student attempted to sexually assault you

The rest: 8.00
PNTA: 0.00
Total: 1.63

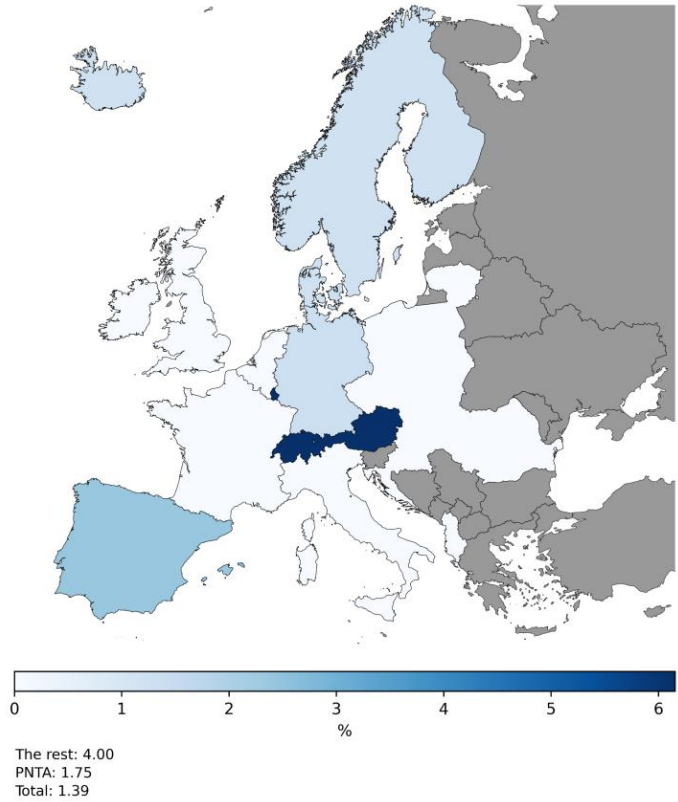

Another economist or economics student sexually assaulted you

# 4. Comparison with other surveys

## 4.1 Comparison with the AEA Professional Climate Survey (2018)

The survey population of the first round of the AEA survey includes any individual who was a member of the AEA as of December 1, 2018 or had been a member in the last nine years, including students, faculty, and those working outside of academia. In total, 10,406 individuals responded to the survey (22.9% response rate). The total number of respondents to the EEA survey instead is 861. Table 18 provides a comparison of the characteristics of the respondents between the two surveys.

In the overall sample, there is a larger share of female respondents in the EEA (46.5%) compared to the AEA (30%). Overall, the number of individuals who identified as nonheterosexual, transgender/non-binary, or both is 570 (5.5%) in the AEA 2018 and 84 (9.8%) in the EEA. Moreover, the number of individuals who identified as transgender or non-binary, or both, is 63 (0.6%) for the AEA and 7 (0.8%) for the EEA. Finally, 9.9% of AEA respondents reported having a health condition that affected their work or studies. This percentage is higher in the EEA (19.98% or 172 individuals). The EEA reports a lower rate of Asian (5.8% vs. 15%) and Black respondents (0.7% vs. 3%) than the AEA. Further distinctions include the generally younger demographic of EEA respondents, with a mean age of 39.5 compared to 47.3 in the AEA and a higher likelihood of being politically left-wing (38.7% vs. 30%). Despite both groups having a high marriage rate among women, EEA women have fewer dependents (42.3% vs. 49%).

It is important to emphasize that the following comparison of the results between the EEA and AEA surveys should be taken with caution, particularly when considering respondents with disabilities, ethnic minorities, and LGBTQ+ individuals. This is because there are some methodological differences in how these groups were identified in the two surveys. Specifically, in the EEA survey, individuals who selected the "prefer not to answer" option for the identification question were included in the corresponding category of respondents, whereas this was not the case in the AEA survey. This distinction may influence the interpretation of the data and the comparability of the results across the two surveys.

Thus, it should be noted that the comparison of the results for LGBTQ+, disability, and ethnic minority should be taken with caution, as the dummies were not constructed in the same way across the two surveys. The main reason for EEA MinE choosing a different definition was the relatively few number of respondents. Due to the small sample size, excluding PNTA was not an option, and in most cases we chose to err on the conservative side and count them as minorities. In other words, the comparison is between the majority group—rather than the minority group—against and the rest.

### 4.1.1 General Climate

There are significant differences in the perception of the general climate within the field of economics between respondents to the European Economic Association (EEA) survey and the American Economic Association (AEA) 2018 one (Table 19 (a) and (b)). Across the board, including among male and female respondents, individuals in the EEA report lower satisfaction with the overall climate and feel they have less power at their institutions compared to their counterparts in the AEA.

Specifically, only 28% of respondents in the EEA feel valued within the field of economics, compared to 40% in the AEA. This difference is particularly noticeable among men, with 36% feeling valued in the EEA versus 46% in the AEA. Respondents with disabilities also report feeling less valued in the EEA (20%) compared to the AEA (30%).

LGBTQ+ respondents and those with disabilities in the EEA consistently report lower levels of satisfaction and inclusion at their institutions. For instance, 46% of LGBTQ+ respondents in the EEA feel valued at their institutions, compared to 58% in the AEA. Similarly, only 40% of respondents with disabilities in the EEA feel valued, versus 52% in the AEA. Additionally, these groups are less likely to feel included intellectually, believe their work is valued, or feel they have significant power at their institutions.

Ethnic minority respondents in the EEA also report a less favorable experience compared to non-white respondents in the AEA. For example, only 19% of ethnic minority respondents in the EEA are satisfied with the overall climate in the field of economics, compared to 34% of non-white respondents in the AEA. They also report lower feelings of being valued, socially included, and having significant power at their institution.

### 4.1.2 Opinions and perceptions

The survey highlights several differences in opinions and perceptions between respondents from the European Economic Association (EEA) and the American Economic Association (AEA) 2018 (Table 20). Across the general, female, and male respondent groups, a smaller percentage of people in the EEA believe that individuals of their sexual orientation or those with disabilities are respected within the field.

Men in the EEA are also less likely to believe that discrimination is rare within economics (8% compared to 16% in the AEA), that the economics discipline would be more vibrant if it were more diverse (50% compared to 60%), and that transgender and gender non-conforming individuals are respected within the field (26% compared 38%).

Among LGBTQ+ individuals, those with disabilities, and ethnic minorities, fewer respondents in the EEA than in the AEA believe that discrimination is rare within the field of economics. However, within the EEA, 42% of ethnic minority respondents feel that people of their ethnicity are respected in the field of economics, compared to 30% of non-white respondents in the AEA. Similarly, 41% of LGBTQ+ respondents in the EEA believe their sexual orientation is respected, compared to 29% in the AEA. Despite these findings, respondents with disabilities in the EEA are less likely to believe that people with a disability are respected within the field of economics compared to their AEA counterparts (28% compared to 39%). This trend is consistent for LGBTQ+ individuals and ethnic minorities, who also report lower levels of perceived respect for people with a disability in the EEA.

### 4.1.3 Direct experiences of discrimination

Notably, in the EEA, a higher percentage of men report experiencing discrimination based on sex relative to the AEA 2018 survey (12% compared to 4%) (Table 21). Women in the EEA are also more likely to have faced discrimination based on their research topics, with 33% reporting such experiences compared to 23% in the AEA.

The data also show that individuals with disabilities or those belonging to ethnic minorities in the EEA are more likely to have encountered sex-based discrimination. However, when it comes to sexual orientation, a higher percentage of LGBTQ+ respondents in the AEA report discrimination (18%) compared to their counterparts in the EEA (7%). Similarly, 14% of respondents with disabilities in the AEA report having faced discrimination based on their disability status, compared to 6% in the EEA.

Additionally, 33% of ethnic minority respondents in the EEA have experienced discrimination related to their research topics, which is higher than the 20% of non-white respondents in the AEA who report similar experiences.

### 4.1.4 Witnessing discrimination

The survey also sheds light on the types and frequencies of discrimination witnessed by respondents within the field of economics (Table 22). Among women in the EEA, 27% report having witnessed discrimination based on sex, which is notably lower than the 44% reported by women in the AEA 2018 survey. Conversely, 29% of men in the EEA have observed discrimination related to place of employment, compared to just 17% in the AEA. It is also worth noting that 15% of EEA respondents have witnessed discrimination based on gender identity (9% among the AEA respondents), a rather striking share given the small share of transgender respondents among the respondents and the general population (around 1%).

Regarding LGBTQ+ respondents in the EEA, 34% have witnessed sex-based discrimination (compared to 46% in the AEA), 14% (compared to 24%) have seen discrimination based on sexual orientation, and 19% (compared to 29%) have observed discrimination related to political views.

Similarly, a smaller percentage of respondents with disabilities in the EEA have witnessed discrimination based on sex (29%) compared to those in the AEA (40%).

### 4.1.5 Experiences of discrimination and unfair treatment while student

The survey findings highlight differences in experiences of discrimination and unfair treatment among students in economics, particularly concerning the job market (Table 23). In the EEA survey, a smaller percentage of women report personally experiencing discrimination in relation to the job market, with 19% indicating such experiences compared to 36% in the AEA 2018. Similarly, 16% of LGBTQ+ respondents in the EEA report job market discrimination, which is also lower than the 28% observed in the AEA.

### 4.1.6 Experiences of discrimination in academia

The survey reveals notable differences in the experiences of discrimination faced by women and other minority groups in academia between the two surveys (Table 24). In the EEA, a smaller percentage of women report experiencing discrimination in several critical areas of academic life. Specifically, 27% report discrimination related to compensation (compared to 37% in the AEA), 29% related to service obligations (compared to 43%), 28% related to course evaluations (compared to 47%), 22% related to publishing decisions (compared to 31%), and 18% related to invitations to participate in research conferences (compared to 32%).

While course evaluation discrimination for women is lower in the EEA than in the AEA, it remains significant, reflecting trends observed in previous research, such as the findings by Boring (2017).

Among LGBTQ+ respondents, 13% report experiencing discrimination related to service obligations in the EEA, compared to 24% in the AEA. Furthermore, across the LGBTQ+, those with a disability, and ethnic minority groups, a smaller percentage of respondents report discrimination based on course evaluations in the EEA compared to the AEA. Nevertheless, between one-quarter and one-third of these minority individuals report having personally experienced discrimination in hiring decisions.

It is also worth noting that hostility during presentations and talks is still widespread, as reported by a significant fraction of EEA female respondents (32%), by individuals with disability (28%), ethnic minorities (25%), and LGBTQ+ respondents (19%).

### 4.1.7 Actions taken to avoid possible harassment, discrimination, or unfair or disrespectful treatment

Across all groups, a higher percentage of respondents in the EEA refrained from speaking at conferences or during seminar presentations to avoid possible harassment (Table 25). This trend is especially pronounced among men (30% in the EEA compared to. 19% in the AEA) and women (56% in the EEA compared to 46% in the AEA).

Among women, 11% in the EEA report having changed the topic, content, or method of a class they teach to avoid possible harassment, compared to 20% in the AEA.

For LGBTQ+ respondents, 21% in the EEA chose not to apply for or accept certain employment positions to avoid potential harassment, which is lower compared to 34% in the AEA.

The survey also highlights that a larger share of respondents with disabilities in the EEA (55% compared to 36%) have avoided speaking at conferences or during seminars. Similarly, a higher percentage of ethnic minority respondents in the EEA (43% compared to 30%) have taken similar actions to avoid potential discriminatory or disrespectful treatment.

**4.1.8 Experiences of Exclusion and Harassment**

In the EEA, 50% of men report having felt socially excluded at meetings or events within the field, compared to 40% of men in the AEA 2018 (Table 26). This trend extends to perceptions of work valuation, with 57% of men in the EEA feeling that their work was not taken as seriously as that of their colleagues, whereas 43% of men in the AEA report similar experiences. The disparity is also significant for women, with 75% in the EEA reporting such an experience compared to 69% in the AEA.

Moreover, 50% of men in the EEA have felt that the subject or methodology of their research was not taken as seriously as that of their colleagues, in contrast to 40% in the AEA. Additionally, a larger percentage of respondents from ethnic minority groups in the EEA report feeling disrespected by their economist colleagues, with 59% expressing this sentiment compared to 45% in the AEA.

Table 27 provides additional information on the experiences of assault across the two surveys. Among those respondents who experienced sexual assault, it is common that this took place at their university, with similar percentages across the EEA (32%) and the AEA (34%). However, differences emerge in other contexts. For instance, sexual assaults at another university were less frequently reported by EEA respondents (5%) compared to AEA respondents (16%). Assaults at professional conferences or meetings were reported by 26% of EEA respondents, a significantly higher percentage than the 12% reported by AEA respondents. Similarly, incidents occurring during a conference but not at the conference venue were more common among EEA respondents (26%) than AEA respondents (9%).

For most of the EEA respondents who experienced assault, the perpetrator was someone with authority, such as a professor (54%), compared to 23% in the AEA. Conversely, a more significant proportion of AEA respondents who had such experience reported being assaulted by another economist or student they knew (45%), compared to 26% in the EEA.

Approximately 37% of respondents who experienced assault in the EEA and 36% for the AEA indicated that they reported the assault. Regarding the reasons for not reporting, both EEA and AEA respondents cited concerns about confidentiality as a significant barrier (37% and 40%, respectively). Moreover, a larger proportion (37%) of AEA respondents reported that they did not want or need any action taken, compared to 16% of EEA respondents.

Finally, respondents from both surveys report a profound impact on their professional lives, with around 50% of AEA respondents who experienced assault considering leaving a project or program as a result of the assault. This percentage is lower (32%) for EEA respondents. Moreover, 67% of AEA respondents reported being less productive or effective in their work following the incident, compared to 53% of EEA respondents.

Table 18: Comparison of survey respondents' characteristics, AEA 2018.

|  | EEA | | AEA 2018 | |
|---|---|---|---|---|
|  | N | Mean | N | Mean |
| Female | 861 | 0.465 | 10,305 | 0.30 |
| White | 861 | 0.765 | 10,247 | 0.79 |
| Black | 861 | 0.007 | 10,247 | 0.03 |
| Asian | 861 | 0.058 | 10,247 | 0.15 |
| LGBTQ+ | 861 | 0.173 | 10,332 | 0.06 |
| Transgender | 861 | 0.001 | 10,128 | 0.00 |
| With some disability | 861 | 0.250 | 10,258 | 0.10 |
| Student | 861 | 0.074 | 10,264 | 0.06 |
| Among employed, employer is: | | | | |
|     College or University | 861 | 0.780 | 9,031 | 0.75 |
|     For-profit organization | 861 | 0.005 | 9,031 | 0.07 |
|     Not-for-profit organization | 861 | 0.005 | 9,031 | 0.05 |
| Age | 861 | 39.45 | 9,974 | 47.32 |
| Married | 861 | 0.725 | 10,246 | 0.80 |
| Jewish | 861 | 0.017 | 10,109 | 0.07 |
| Muslim | 861 | 0.016 | 10,109 | 0.03 |
| Among women: | | | | |
|     Married | 400 | 0.693 | 3,042 | 0.75 |
|     With dependents | 400 | 0.423 | 3,053 | 0.49 |
|     Employer is a college or university | 400 | 0.770 | 2,706 | 0.78 |

Table 19: General climate, comparison with AEA 2018.

(a) All, male, and female

|  | EEA | | | AEA 2018 | | |
|---|---|---|---|---|---|---|
|  | All | Male | Female | All | Male | Female |
| I am satisfied with the overall climate within the field of economics | 0.23 | 0.31 | 0.15 | 0.34 | 0.40 | 0.20 |
| I am satisfied with the overall climate at my institution/place of employment | 0.54 | 0.63 | 0.45 | 0.56 | 0.61 | 0.46 |
| I feel valued within the field of economics | 0.28 | 0.36 | 0.21 | 0.40 | 0.46 | 0.25 |
| I feel valued at my institution/place of employment | 0.54 | 0.62 | 0.47 | 0.63 | 0.67 | 0.54 |
| I always feel included socially within the field of economics | 0.31 | 0.40 | 0.22 | 0.37 | 0.43 | 0.21 |
| I always feel included socially at my institution/place of employment | 0.57 | 0.67 | 0.49 | 0.63 | 0.67 | 0.52 |
| I always feel included intellectually within the field of economics | 0.37 | 0.46 | 0.29 | 0.42 | 0.48 | 0.27 |
| I always feel included intellectually at my institution/place of employment | 0.55 | 0.63 | 0.46 | 0.64 | 0.69 | 0.53 |
| I feel I have been discriminated against within the field of economics | 0.22 | 0.19 | 0.25 | 0.18 | 0.13 | 0.30 |
| I feel I have been discriminated against at my institution/place of employment | 0.13 | 0.08 | 0.16 | 0.12 | 0.09 | 0.20 |
| The work that I do is valued within the field of economics | 0.36 | 0.39 | 0.32 | 0.40 | 0.43 | 0.33 |
| The work that I do is valued at my institution/place of employment | 0.54 | 0.60 | 0.49 | 0.63 | 0.65 | 0.56 |
| I think I have a great deal of power within the field of economics | 0.04 | 0.05 | 0.03 | 0.07 | 0.08 | 0.05 |
| I think I have a great deal of power at my institution/place of employment | 0.12 | 0.15 | 0.09 | 0.24 | 0.25 | 0.20 |
| My ideas and opinions are often ignored within the field of economics | 0.16 | 0.16 | 0.14 | 0.21 | 0.20 | 0.23 |
| My ideas and opinions are often ignored at my institution/place of employment | 0.12 | 0.10 | 0.13 | 0.13 | 0.12 | 0.14 |

Reported in each cell is the share of respondents who agree or strongly agree with each statement

Table 19: General climate, comparison with AEA 2018.
(b) LGBTQ+, with disability, and ethnic minority

|  | EEA | | | AEA 2018 | | |
| --- | --- | --- | --- | --- | --- | --- |
|  | LGBTQ+ | With disability | Ethnic minority | LGBTQ+ | With disability | Non-white |
| I am satisfied with the overall climate within the field of economics | 0.21 | 0.21 | 0.19 | 0.26 | 0.26 | 0.34 |
| I am satisfied with the overall climate at my institution/place of employment | 0.45 | 0.40 | 0.46 | 0.52 | 0.45 | 0.50 |
| I feel valued within the field of economics | 0.26 | 0.20 | 0.22 | 0.34 | 0.30 | 0.36 |
| I feel valued at my institution/place of employment | 0.46 | 0.40 | 0.48 | 0.58 | 0.52 | 0.55 |
| I always feel included socially within the field of economics | 0.23 | 0.24 | 0.22 | 0.28 | 0.29 | 0.32 |
| I always feel included socially at my institution/place of employment | 0.49 | 0.50 | 0.50 | 0.53 | 0.52 | 0.56 |
| I always feel included intellectually within the field of economics | 0.34 | 0.27 | 0.33 | 0.35 | 0.35 | 0.40 |
| I always feel included intellectually at my institution/place of employment | 0.47 | 0.42 | 0.51 | 0.58 | 0.53 | 0.59 |
| I feel I have been discriminated against within the field of economics | 0.26 | 0.27 | 0.31 | 0.27 | 0.25 | 0.25 |
| I feel I have been discriminated against at my institution/place of employment | 0.14 | 0.21 | 0.17 | 0.16 | 0.17 | 0.16 |
| The work that I do is valued within the field of economics | 0.29 | 0.29 | 0.34 | 0.35 | 0.33 | 0.41 |
| The work that I do is valued at my institution/place of employment | 0.46 | 0.43 | 0.51 | 0.58 | 0.53 | 0.58 |
| I think I have a great deal of power within the field of economics | 0.03 | 0.02 | 0.05 | 0.08 | 0.06 | 0.11 |
| I think I have a great deal of power at my institution/place of employment | 0.07 | 0.07 | 0.10 | 0.19 | 0.18 | 0.22 |
| My ideas and opinions are often ignored within the field of economics | 0.21 | 0.21 | 0.22 | 0.28 | 0.29 | 0.24 |
| My ideas and opinions are often ignored at my institution/place of employment | 0.13 | 0.15 | 0.17 | 0.13 | 0.19 | 0.15 |

Reported in each cell is the share of respondents who agree or strongly agree with each statement

Table 20: Opinions and perceptions, comparison with AEA 2018.

(a) All, male, and female

|  | EEA | | | AEA 2018 | | |
|---|---|---|---|---|---|---|
|  | All | Male | Female | All | Male | Female |
| It is not important for the field of economics to be inclusive towards people with different backgrounds. | 0.11 | 0.12 | 0.09 | 0.07 | 0.08 | 0.05 |
| Discrimination is rare within the field of economics. | 0.06 | 0.08 | 0.03 | 0.12 | 0.16 | 0.04 |
| Economics would be a more vibrant discipline if it were more diverse. | 0.66 | 0.50 | 0.83 | 0.66 | 0.60 | 0.82 |
| People of my ethnicity are respected within the field of economics | 0.74 | 0.76 | 0.75 | 0.71 | 0.73 | 0.67 |
| People who are not White are respected within the field of economics | 0.36 | 0.48 | 0.24 | 0.45 | 0.51 | 0.28 |
| Men are respected within the field of economics | 0.86 | 0.79 | 0.96 | 0.90 | 0.88 | 0.96 |
| Women are respected within the field of economics | 0.33 | 0.51 | 0.13 | 0.40 | 0.51 | 0.16 |
| Transgender and gender non-conforming people are respected within the field of economics | 0.20 | 0.26 | 0.13 | 0.32 | 0.38 | 0.16 |
| People of my sexual orientation are respected within the field of economics | 0.72 | 0.78 | 0.67 | 0.87 | 0.88 | 0.86 |
| People who are not heterosexual are respected within the field of economics | 0.46 | 0.58 | 0.33 | 0.57 | 0.62 | 0.42 |
| People of my disability status are respected within the field |  |  |  | 0.52 | 0.58 | 0.36 |
| People with a disability are respected within the field of economics | 0.33 | 0.46 | 0.18 | 0.58 | 0.63 | 0.40 |
| At your university, the discussion culture during academic presentations is productive and fair | 0.54 | 0.63 | 0.45 |  |  |  |
| At international conferences in Europe, the discussion culture during academic presentations is productive and fair | 0.42 | 0.48 | 0.35 |  |  |  |

Reported in each cell is the share of respondents who agree or strongly agree with each statement.

Table 20: Opinions and perceptions, comparison with AEA 2018.
(b) LGBTQ+, with disability, and ethnic minority

|  | EEA | | | AEA 2018 | | |
|---|---|---|---|---|---|---|
|  | LGBTQ+ | With disability | Ethnic minority | LGBTQ+ | With disability | Non-white |
| It is not important for the field of economics to be inclusive towards people with different backgrounds. | 0.13 | 0.12 | 0.14 | 0.08 | 0.08 | 0.09 |
| Discrimination is rare within the field of economics. | 0.04 | 0.06 | 0.05 | 0.12 | 0.13 | 0.11 |
| Economics would be a more vibrant discipline if it were more diverse. | 0.68 | 0.67 | 0.65 | 0.77 | 0.67 | 0.73 |
| People of my ethnicity are respected within the field of economics | 0.63 | 0.67 | 0.42 | 0.65 | 0.69 | 0.30 |
| People who are not White are respected within the field of economics | 0.29 | 0.32 | 0.31 | 0.34 | 0.42 | 0.28 |
| Men are respected within the field of economics | 0.81 | 0.84 | 0.83 | 0.92 | 0.89 | 0.78 |
| Women are respected within the field of economics | 0.30 | 0.32 | 0.31 | 0.30 | 0.39 | 0.30 |
| Transgender and gender non-conforming people are respected within the field of economics | 0.13 | 0.21 | 0.21 | 0.18 | 0.28 | 0.22 |
| People of my sexual orientation are respected within the field of economics | 0.41 | 0.60 | 0.59 | 0.29 | 0.84 | 0.74 |
| People who are not heterosexual are respected within the field of economics | 0.29 | 0.42 | 0.39 | 0.28 | 0.51 | 0.40 |
| People of my disability status are respected within the field |  |  |  | 0.38 | 0.34 | 0.42 |
| People with a disability are respected within the field of economics | 0.27 | 0.28 | 0.29 | 0.39 | 0.39 | 0.45 |
| At your university, the discussion culture during academic presentations is productive and fair | 0.51 | 0.47 | 0.49 |  |  |  |
| At international conferences in Europe, the discussion culture during academic presentations is productive and fair | 0.33 | 0.38 | 0.34 |  |  |  |

Reported in each cell is the share of respondents who agree or strongly agree with each statement.

Table 21: Directly experienced discrimination - by type of discrimination, comparison with AEA 2018.

(a) All, male, and female

|  | EEA | | | AEA 2018 | | |
|---|---|---|---|---|---|---|
|  | All | Male | Female | All | Male | Female |
| Have you ever been discriminated against, or treated unfairly based on: | | | | | | |
| Racial/ethnic identity | 0.07 | 0.07 | 0.07 | 0.09 | 0.08 | 0.12 |
| Sex | 0.30 | 0.12 | 0.49 | 0.17 | 0.04 | 0.48 |
| Gender identity | 0.07 | 0.04 | 0.11 | | | |
| Sexual orientation | 0.02 | 0.02 | 0.01 | 0.02 | 0.02 | 0.02 |
| Disability status | 0.02 | 0.01 | 0.02 | 0.02 | 0.01 | 0.03 |
| Marital status/caregiving responsibilities | 0.13 | 0.06 | 0.20 | 0.09 | 0.04 | 0.22 |
| Religion | 0.02 | 0.03 | 0.01 | 0.05 | 0.05 | 0.05 |
| Political views | 0.09 | 0.09 | 0.08 | 0.09 | 0.09 | 0.10 |
| Age | 0.14 | 0.10 | 0.18 | 0.09 | 0.06 | 0.16 |
| Citizenship status | 0.10 | 0.09 | 0.11 | 0.06 | 0.06 | 0.09 |
| Place of employment | 0.23 | 0.20 | 0.27 | 0.15 | 0.13 | 0.22 |
| Research topics | 0.27 | 0.20 | 0.33 | 0.16 | 0.13 | 0.23 |
| Other factors | | | | 0.09 | 0.07 | 0.13 |

Reported in each cell is the share of respondents that report having personally experienced these different types of discrimination or unfair treatment over the last ten years in the field of economics.

Table 21: Directly experienced discrimination - by type of discrimination, comparison with AEA 2018.

(b) LGBTQ+, with disability, and ethnic minority

|  | EEA | | | AEA 2018 | | |
|---|---|---|---|---|---|---|
|  | LGBTQ+ | With disability | Ethnic minority | LGBTQ+ | With disability | Non-white |
| Have you ever been discriminated against, or treated unfairly based on: | | | | | | |
| Racial/ethnic identity | 0.09 | 0.08 | 0.20 | 0.11 | 0.12 | 0.29 |
| Sex | 0.31 | 0.34 | 0.27 | 0.23 | 0.23 | 0.17 |
| Gender identity | 0.08 | 0.08 | 0.08 | | | |
| Sexual orientation | 0.07 | 0.02 | 0.03 | 0.18 | 0.03 | 0.02 |
| Disability status | 0.03 | 0.06 | 0.03 | 0.05 | 0.14 | 0.02 |
| Marital status / caregiving responsibilities | 0.10 | 0.14 | 0.14 | 0.11 | 0.13 | 0.10 |
| Religion | 0.01 | 0.02 | 0.04 | 0.05 | 0.08 | 0.07 |
| Political views | 0.12 | 0.11 | 0.14 | 0.12 | 0.14 | 0.10 |
| Age | 0.13 | 0.15 | 0.16 | 0.09 | 0.16 | 0.12 |
| Citizenship status | 0.15 | 0.14 | 0.19 | 0.09 | 0.08 | 0.16 |
| Place of employment | 0.23 | 0.26 | 0.27 | 0.18 | 0.20 | 0.20 |
| Research topics | 0.32 | 0.30 | 0.33 | 0.24 | 0.23 | 0.20 |
| Other factors | | | | 0.13 | 0.14 | 0.14 |

Reported in each cell is the share of respondents that report having personally experienced these different types of discrimination or unfair treatment over the last ten years in the field of economics.

Table 22: Witnessed discrimination - by type of discrimination, comparison with AEA 2018.

(a) All, male, and female

|  | EEA | | | AEA 2018 | | |
|---|---|---|---|---|---|---|
|  | All | Male | Female | All | Male | Female |
| Have you ever witnessed discrimination or unfair treatment based on: | | | | | | |
| Racial/ethnic identity | 0.21 | 0.15 | 0.27 | 0.23 | 0.19 | 0.33 |
| Sex | 0.33 | 0.38 | 0.27 | 0.37 | 0.33 | 0.44 |
| Gender identity | 0.15 | 0.11 | 0.18 | 0.09 | 0.08 | 0.12 |
| Sexual orientation | 0.08 | 0.06 | 0.11 | 0.09 | 0.08 | 0.13 |
| Disability status | 0.06 | 0.05 | 0.07 | 0.06 | 0.05 | 0.10 |
| Marital status/caregiving responsibilities | 0.25 | 0.18 | 0.33 | 0.18 | 0.13 | 0.31 |
| Religion | 0.07 | 0.06 | 0.07 | 0.09 | 0.09 | 0.10 |
| Political views | 0.20 | 0.22 | 0.20 | 0.23 | 0.22 | 0.22 |
| Age | 0.21 | 0.17 | 0.24 | 0.17 | 0.15 | 0.22 |
| Citizenship status | 0.18 | 0.15 | 0.21 | 0.15 | 0.13 | 0.19 |
| Place of employment | 0.29 | 0.29 | 0.29 | 0.19 | 0.17 | 0.24 |
| Research topics | 0.32 | 0.29 | 0.33 | 0.23 | 0.22 | 0.27 |
| Other factors | | | | 0.09 | 0.09 | 0.11 |

Reported in each cell is the share of respondents who have witnessed different types of discrimination or unfair treatment over the last ten years in the field of economics.

Table 22: Witnessed discrimination - by type of discrimination, comparison with AEA 2018.
(b) LGBTQ+, with disability, and ethnic minority

|  | EEA | | | AEA 2018 | | |
|---|---|---|---|---|---|---|
|  | LGBTQ+ | With disability | Ethnic minority | LGBTQ+ | With disability | Non-white |
| Have you ever witnessed discrimination or unfair treatment based on: | | | | | | |
| Racial/ethnic identity | 0.26 | 0.27 | 0.21 | 0.35 | 0.29 | 0.28 |
| Sex | 0.34 | 0.29 | 0.32 | 0.46 | 0.40 | 0.30 |
| Gender identity | 0.24 | 0.22 | 0.16 | 0.19 | 0.11 | 0.10 |
| Sexual orientation | 0.14 | 0.09 | 0.08 | 0.24 | 0.13 | 0.10 |
| Disability status | 0.09 | 0.09 | 0.06 | 0.11 | 0.14 | 0.08 |
| Marital status / caregiving responsibilities | 0.31 | 0.30 | 0.23 | 0.22 | 0.23 | 0.18 |
| Religion | 0.08 | 0.11 | 0.08 | 0.13 | 0.12 | 0.12 |
| Political views | 0.19 | 0.26 | 0.19 | 0.29 | 0.26 | 0.22 |
| Age | 0.23 | 0.20 | 0.19 | 0.19 | 0.18 | 0.16 |
| Citizenship status | 0.16 | 0.17 | 0.17 | 0.21 | 0.20 | 0.20 |
| Place of employment | 0.27 | 0.28 | 0.28 | 0.26 | 0.23 | 0.20 |
| Research topics | 0.32 | 0.34 | 0.27 | 0.28 | 0.29 | 0.21 |
| Other factors | | | | 0.14 | 0.13 | 0.12 |

Reported in each cell is the share of respondents who have witnessed different types of discrimination or unfair treatment over the last ten years in the field of economics.

Table 23: Experiences of discrimination and unfair treatment while student, comparison with AEA 2018.

(a) All, male, and female

|  | EEA | | | AEA 2018 | | |
| --- | --- | --- | --- | --- | --- | --- |
|  | All | Male | Female | All | Male | Female |
| During your time as a student studying economics, have you personally experienced discrimination or unfair treatment concerning: | | | | | | |
| Access to research assistantships | 0.12 | 0.10 | 0.15 | 0.10 | 0.07 | 0.18 |
| Access to advisors | 0.12 | 0.09 | 0.15 | 0.10 | 0.06 | 0.20 |
| Access to quality advising | 0.15 | 0.10 | 0.21 | 0.12 | 0.07 | 0.23 |
| Job market | 0.19 | 0.19 | 0.19 | 0.21 | 0.15 | 0.36 |

Reported in each cell is the share of respondents who report having personally experienced these treatments during their time studying economics.

Table 23: Experiences of discrimination and unfair treatment while student, comparison with AEA 2018.

(b) LGBTQ+, with disability, and ethnic minority

|  | EEA | | | AEA 2018 | | |
|---|---|---|---|---|---|---|
|  | LGBTQ+ | With disability | Ethnic minority | LGBTQ+ | With disability | Non-white |
| During your time as a student studying economics, have you personally experienced discrimination or unfair treatment concerning: | | | | | | |
| Access to research assistantships | 0.13 | 0.16 | 0.17 | 0.15 | 0.15 | 0.18 |
| Access to advisors | 0.11 | 0.15 | 0.13 | 0.15 | 0.15 | 0.17 |
| Access to quality advising | 0.18 | 0.20 | 0.18 | 0.17 | 0.18 | 0.19 |
| Job market | 0.16 | 0.22 | 0.23 | 0.28 | 0.30 | 0.31 |

Reported in each cell is the share of respondents who report having personally experienced these treatments during their time studying economics.

Table 24: Experiences of discrimination in academia, comparison with AEA 2018.

(a) All, male, and female

|  | EEA | | | AEA 2018 | | |
|---|---|---|---|---|---|---|
|  | All | Male | Female | All | Male | Female |
| Have you personally experienced discrimination or unfair treatment with regard to: | | | | | | |
| Hiring decisions | 0.24 | 0.25 | 0.23 | | | |
| Promotion decisions | 0.19 | 0.13 | 0.24 | 0.16 | 0.11 | 0.27 |
| Compensation | 0.19 | 0.12 | 0.27 | 0.20 | 0.12 | 0.37 |
| Teaching assignments | 0.15 | 0.08 | 0.21 | 0.15 | 0.09 | 0.28 |
| Service obligations | 0.18 | 0.07 | 0.29 | 0.19 | 0.09 | 0.43 |
| Committee or decision-making body membership | 0.17 | 0.10 | 0.24 | | | |
| Access to time and funding to attend conferences and seminars | 0.12 | 0.10 | 0.14 | 0.10 | 0.07 | 0.16 |
| Access to graduate student researchers or research assistants | 0.11 | 0.08 | 0.14 | 0.08 | 0.05 | 0.15 |
| Course evaluations | 0.16 | 0.05 | 0.28 | 0.20 | 0.08 | 0.47 |
| Publishing decisions | 0.20 | 0.18 | 0.22 | 0.22 | 0.18 | 0.31 |
| Funding decisions | 0.14 | 0.13 | 0.15 | 0.14 | 0.12 | 0.20 |
| Sabbatical time | 0.04 | 0.03 | 0.05 | 0.05 | 0.04 | 0.09 |
| Access to potential coauthors | 0.11 | 0.05 | 0.18 | 0.11 | 0.06 | 0.21 |
| Invitations to participate in reserach conferences, associations and networks | 0.18 | 0.18 | 0.18 | 0.19 | 0.14 | 0.32 |
| Seminar invitations | 0.15 | 0.14 | 0.15 | | | |
| Hostility during presentations/talks | 0.23 | 0.16 | 0.32 | | | |

Reported in each cell is the share of respondents who report having personally experienced discrimination or unfair treatment in these aspects of work over the last ten years in the field of economics.

Table 24: Experiences of discrimination in academia, comparison with AEA 2018.

(b) LGBTQ+, with disability, and ethnic minority

|  | EEA | | | AEA 2018 | | |
| --- | --- | --- | --- | --- | --- | --- |
|  | LGBTQ+ | With disability | Ethnic minority | LGBTQ+ | With disability | Non-white |
| Have you personally experienced discrimination or unfair treatment with regard to: | | | | | | |
| Hiring decisions | 0.25 | 0.28 | 0.31 | | | |
| Promotion decisions | 0.15 | 0.21 | 0.23 | 0.21 | 0.22 | 0.22 |
| Compensation | 0.17 | 0.23 | 0.19 | 0.22 | 0.24 | 0.26 |
| Teaching assignments | 0.14 | 0.16 | 0.17 | 0.18 | 0.20 | 0.21 |
| Service obligations | 0.13 | 0.20 | 0.18 | 0.24 | 0.25 | 0.24 |
| Committee or decision-making body membership | 0.14 | 0.20 | 0.17 | | | |
| Access to time and funding to attend conferences and seminars | 0.09 | 0.14 | 0.16 | 0.15 | 0.14 | 0.16 |
| Access to graduate student researchers or research assistants | 0.09 | 0.13 | 0.15 | 0.10 | 0.13 | 0.13 |
| Course evaluations | 0.17 | 0.20 | 0.14 | 0.28 | 0.30 | 0.27 |
| Publishing decisions | 0.18 | 0.23 | 0.23 | 0.23 | 0.27 | 0.29 |
| Funding decisions | 0.12 | 0.18 | 0.17 | 0.17 | 0.22 | 0.20 |
| Sabbatical time | 0.02 | 0.04 | 0.05 | 0.07 | 0.09 | 0.08 |
| Access to potential coauthors | 0.11 | 0.17 | 0.16 | 0.14 | 0.16 | 0.16 |
| Invitations to participate in research conferences, associations and networks | 0.18 | 0.21 | 0.24 | 0.22 | 0.26 | 0.27 |
| Seminar invitations | 0.10 | 0.16 | 0.21 | | | |
| Hostility during presentations/talks | 0.19 | 0.28 | 0.25 | | | |

Reported in each cell is the share of respondents who report having personally experienced discrimination or unfair treatment in these aspects of work over the last ten years in the field of economics.

Table 25: Actions taken to avoid possible harassment, discrimination, or unfair treatment, comparison with AEA 2018.

(a) All, male, and female

|  | EEA | | | AEA 2018 | | |
|---|---|---|---|---|---|---|
|  | All | Male | Female | All | Male | Female |
| Have you ever done any of the following to avoid possible harassment, discrimination, or unfair or disrespectful treatment: | | | | | | |
| Not applied for or accepted admission at a particular grad school | 0.09 | 0.06 | 0.11 | 0.07 | 0.05 | 0.10 |
| Paused or ceased enrollment at a particular grad school | 0.03 | 0.02 | 0.04 | 0.02 | 0.02 | 0.04 |
| Not applied for or taken a particular employment position | 0.18 | 0.12 | 0.22 | 0.16 | 0.12 | 0.24 |
| Not applied for or taken a promotion at your place of employment | 0.08 | 0.06 | 0.10 | 0.07 | 0.05 | 0.12 |
| Left a particular employment position | 0.13 | 0.09 | 0.17 | 0.10 | 0.08 | 0.15 |
| Not presented your question, idea, or view at your school or place of work | 0.40 | 0.26 | 0.54 | 0.31 | 0.24 | 0.47 |
| Not participated in a conference | 0.20 | 0.16 | 0.24 | 0.14 | 0.11 | 0.21 |
| Not spoken at a conference or during a seminar presentation | 0.42 | 0.30 | 0.56 | 0.27 | 0.19 | 0.46 |
| Not made a professional visit to a particular place | 0.13 | 0.09 | 0.17 | 0.11 | 0.08 | 0.16 |
| Not attended social events after class, at work, or at conferences | 0.30 | 0.17 | 0.43 | 0.25 | 0.18 | 0.43 |
| Changed the topic, content, or method of a class you teach | 0.10 | 0.09 | 0.11 | 0.14 | 0.12 | 0.20 |
| Changed the content, method, or conclusions of a research paper | 0.14 | 0.12 | 0.15 | 0.09 | 0.08 | 0.10 |
| Not started or continued research in a particular field | 0.23 | 0.18 | 0.28 | 0.17 | 0.14 | 0.25 |

Reported in each cell is the share of respondents who report having taken the listed action over the last ten years in the field of economics.

Table 25: Actions taken to avoid possible harassment, discrimination, or unfair treatment, comparison with AEA 2018.
(b) LGBTQ+, with disability, and ethnic minority

|  | EEA | | | AEA 2018 | | |
|---|---|---|---|---|---|---|
|  | LGBTQ+ | With disability | Ethnic minority | LGBTQ+ | With disability | Non-white |
| Have you ever done any of the following to avoid possible harassment, discrimination, or unfair or disrespectful treatment: | | | | | | |
| Not applied for or accepted admission at a particular grad school | 0.15 | 0.14 | 0.11 | 0.16 | 0.14 | 0.13 |
| Paused or ceased enrollment at a particular grad school | 0.03 | 0.03 | 0.04 | 0.04 | 0.05 | 0.06 |
| Not applied for or taken a particular employment position | 0.21 | 0.20 | 0.19 | 0.34 | 0.24 | 0.23 |
| Not applied for or taken a promotion at your place of employment | 0.07 | 0.08 | 0.13 | 0.10 | 0.10 | 0.12 |
| Left a particular employment position | 0.15 | 0.19 | 0.17 | 0.15 | 0.15 | 0.15 |
| Not presented your question, idea, or view at your school or place of work | 0.40 | 0.48 | 0.43 | 0.44 | 0.45 | 0.36 |
| Not participated in a conference | 0.19 | 0.24 | 0.23 | 0.20 | 0.23 | 0.20 |
| Not spoken at a conference or during a seminar presentation | 0.39 | 0.55 | 0.43 | 0.38 | 0.36 | 0.30 |
| Not made a professional visit to a particular place | 0.13 | 0.22 | 0.13 | 0.18 | 0.16 | 0.17 |
| Not attended social events after class, at work, or at conferences | 0.32 | 0.41 | 0.35 | 0.38 | 0.39 | 0.34 |
| Changed the topic, content, or method of a class you teach | 0.10 | 0.13 | 0.13 | 0.16 | 0.21 | 0.19 |
| Changed the content, method, or conclusions of a research paper | 0.17 | 0.18 | 0.19 | 0.14 | 0.14 | 0.13 |
| Not started or continued research in a particular field | 0.28 | 0.33 | 0.34 | 0.27 | 0.27 | 0.22 |

Reported in each cell is the share of respondents who report having taken the listed action over the last ten years in the field of economics.

Table 26: Experiences of exclusion and harassment, comparison with AEA 2018.

(a) All, male, and female

|  | EEA | | | AEA 2018 | | |
|---|---|---|---|---|---|---|
|  | All | Male | Female | All | Male | Female |
| Have you ever experienced any of the following: | | | | | | |
| Felt socially excluded at a meeting or event in the field | 0.59 | 0.50 | 0.69 | 0.47 | 0.40 | 0.65 |
| Felt disrespected by your economist colleagues | 0.56 | 0.44 | 0.69 | 0.45 | 0.38 | 0.62 |
| Felt that your work was not taken as seriously as that of your economist colleagues | 0.65 | 0.57 | 0.75 | 0.51 | 0.43 | 0.69 |
| Felt that the subject or methodology of your research was not taken as seriously as that of your economist colleagues | 0.56 | 0.50 | 0.64 | 0.46 | 0.40 | 0.59 |
| Another economist or economics student displayed inappropriate behavior towards you, including offensive sexual remarks, remarks about your appearance OR body, inappropriate gestures | 0.26 | 0.10 | 0.42 | 0.22 | 0.13 | 0.43 |
| Another economist or economics student made unwanted attempts to establish a dating, romantic, or sexual relationship with you despite your efforts to discourage it | 0.11 | 0.03 | 0.20 | 0.09 | 0.03 | 0.23 |
| Another economist or economics student threatened you with retaliation for not being romantically or sexually cooperative, or implied you'd be treated better if you were sexually cooperative. | 0.02 | 0.00 | 0.04 | 0.03 | 0.01 | 0.08 |
| Another economist or economics student stalked you | 0.06 | 0.05 | 0.07 | 0.05 | 0.02 | 0.10 |
| Another economist or economics student attempted to sexually assault you | 0.02 | 0.00 | 0.03 | 0.02 | 0.01 | 0.06 |
| Another economist or economics student sexually assaulted you | 0.01 | 0.00 | 0.03 | 0.01 | 0.00 | 0.03 |

Reported in each cell is the share of respondents that report having personally experienced the stated treatment over the last ten years in the field of economics.

Table 26: Experiences of exclusion and harassment, comparison with AEA 2018.

(b) LGBTQ+, with disability, and ethnic minority

|  | EEA | | | AEA 2018 | | |
|---|---|---|---|---|---|---|
|  | LGBTQ+ | With disability | Ethnic minority | LGBTQ+ | With disability | Non-white |
| Have you ever experienced any of the following: | | | | | | |
| Felt socially excluded at a meeting or event in the field | 0.61 | 0.68 | 0.61 | 0.59 | 0.60 | 0.53 |
| Felt disrespected by your economist colleagues | 0.54 | 0.62 | 0.59 | 0.56 | 0.58 | 0.45 |
| Felt that your work was not taken as seriously as that of your economist colleagues | 0.60 | 0.68 | 0.59 | 0.60 | 0.60 | 0.51 |
| Felt that the subject or methodology of your research was not taken as seriously as that of your economist colleagues | 0.52 | 0.62 | 0.53 | 0.54 | 0.55 | 0.46 |
| Another economist or economics student displayed inappropriate behavior towards you, including offensive sexual remarks, remarks about your appearance OR body, inappropriate gestures | 0.26 | 0.36 | 0.27 | 0.37 | 0.29 | 0.20 |
| Another economist or economics student made unwanted attempts to establish a dating, romantic, or sexual relationship with you despite your efforts to discourage it | 0.11 | 0.13 | 0.10 | 0.14 | 0.12 | 0.09 |
| Another economist or economics student threatened you with retaliation for not being romantically or sexually cooperative, or implied you'd be treated better if you were sexually cooperative. | 0.01 | 0.03 | 0.02 | 0.06 | 0.05 | 0.04 |
| Another economist or economics student stalked you | 0.05 | 0.05 | 0.07 | 0.09 | 0.07 | 0.05 |
| Another economist or economics student attempted to sexually assault you | 0.02 | 0.03 | 0.03 | 0.05 | 0.04 | 0.03 |
| Another economist or economics student sexually assaulted you | 0.03 | 0.02 | 0.03 | 0.04 | 0.03 | 0.02 |

Reported in each cell is the share of respondents that report having personally experienced the stated treatment over the last ten years in the field of economics.

Table 27: Experiences of sexual assault.

| | EEA Perc. | EEA N | AEA 2018 Perc. | AEA 2018 N |
|---|---|---|---|---|
| **Where?** | | | | |
| At my university | 0.32 | 6 | 0.34 | 36 |
| At another university | 0.05 | 1 | 0.16 | 17 |
| At my non-academic workplace | 0.05 | 1 | 0.05 | 5 |
| At a professional conference or meeting | 0.26 | 5 | 0.12 | 13 |
| During a conference or meeting, but not at the conference or meeting itself | 0.26 | 5 | 0.09 | 10 |
| Online (via email or other electronic media) | 0.00 | 0 | 0.01 | 1 |
| Somewhere else | 0.00 | 0 | 0.19 | 20 |
| **Who?** | | | | |
| My professor/boss/someone with authority | 0.53 | 10 | 0.23 | 25 |
| A co-worker at my institution or place of employment | 0.11 | 2 | 0.17 | 18 |
| Another economist or student that I know | 0.26 | 5 | 0.45 | 48 |
| Another economist or student that I do not know | 0.05 | 1 | 0.03 | 3 |
| Someone else | 0.00 | 0 | 0.03 | 3 |
| Do not know the identity or status of this person | 0.00 | 0 | 0.03 | 3 |
| **Told anyone?** | | | | |
| Yes | 0.37 | 7 | 0.36 | 39 |
| No | 0.63 | 12 | 0.58 | 62 |
| **Who did you tell first?** | | | | |
| A colleague | 0.57 | 4 | 0.38 | 15 |
| A friend of family member not associated with the field | | | 0.36 | 14 |
| A relevant administration official at the workplace/university/institution | 0.14 | 1 | 0.08 | 3 |
| The police | 0.00 | 0 | 0.00 | 0 |
| Someone else | 0.29 | 2 | 0.18 | 7 |
| **If you did not make a report to some authority, why not?** | | N=19 | | N=60 |
| Didn't know who the right person was | 0.21 | 4 | 0.15 | 9 |
| Were concerned the situation would not be kept confidential | 0.37 | 7 | 0.40 | 24 |
| Did not need/want any assistance or any action taken | 0.16 | 3 | 0.37 | 22 |
| Were concerned the process and/or the outcome would be too difficult | 0.21 | 4 | 0.33 | 20 |
| Were concerned about retribution from the person who did this and/or others over reporting | 0.32 | 6 | 0.40 | 24 |
| Did not think that the person who did this to you will be held accountable | 0.53 | 10 | | |
| Thought the consequences for that person would be too harsh relative to the extent of the misconduct | 0.16 | 3 | | |
| **Did the experience lead you to…** | | N=19 | | N=82 |
| File an official charge of complaint | 0.11 | 2 | 0.05 | 4 |
| File an official charge with police | 0.00 | 0 | 0.05 | 4 |
| Consider leaving a project, committee, program | 0.32 | 6 | 0.50 | 41 |
| Be less productive or effective in your work | 0.53 | 10 | 0.67 | 55 |
| Consider leaving your position | 0.26 | 5 | 0.38 | 31 |
| Take leave, sick time, miss work unexpectedly, or other similar time away from work | 0.11 | 2 | 0.29 | 24 |
| Consider not attending future conferences | 0.32 | 6 | 0.18 | 15 |
| Consider leaving the field of economic research | 0.32 | 6 | 0.23 | 19 |
| Consider leaving academia entirely | 0.32 | 6 | 0.33 | 27 |
| Consider taking legal action | 0.16 | 3 | 0.18 | 15 |

## 4.2 Comparison with Eurobarometer

The "Discrimination in the European Union" (2023) module is part of the special modules of the Eurobarometer ([link](#)). The results refer to all EU countries. Table 14 below shows the percentage of respondents that, where they live, they consider themselves to be part of any minority group. The table also reports the percentage of respondents to the EEA survey that belong to each group (when available).

Table 28: (QSD1) Where you live, do you consider yourself to be part of any of the following?

|  | Eurobarometer | EEA |
|---|---|---|
| Total "Part of a minority group" | 0.11 | |
| A person with a disability | 0.03 | 0.20 |
| A minority in terms of skin color | 0.02 | |
| A religious minority | 0.02 | |
| An ethnic minority | 0.02 | |
| Any other minority group | 0.02 | |
| Being lesbian, gay, or bisexual | 0.02 | 0.09 |
| Being Roma | 0.01 | |

Table 15 below compares the percentage of people who have directly experienced discrimination based on different factors between this survey and the EEA. The questions, however, are slightly different. In the EEA, in particular, there are two questions related to this topic. The first one (Q321) asks respondents whether they have experienced discrimination by anyone in the field of economics in the last 10 years in the field of economics; the second one (Q322) instead is about whether they have been discriminated against by anyone in the field of economics in their current job. The question in the Eurobarometer (QB2) instead asks respondents whether, in the past 12 months, they have personally been discriminated against or have experienced harassment based on different factors. In general, the results from the Eurobarometer are lower but still comparable to the EEA (especially for sexual orientation 1−2%, religion 2%, and disability status 2%). The only factor where there is a very significant difference is sex: while in the EEA the percentage of respondents that have been discriminated against based on sex is 29.5% (Q321) or 18.4% (Q322), the percentage is 5% for the Eurobarometer. Overall, around 21% of respondents (24% of women and 18% of men) say they have personally felt discriminated against or experienced harassment in the past 12 months according to the Eurobarometer.

Table 29: Direct experiences of discrimination, comparison with Eurobarometer

|  | EEA (last 10 years) | EEA (current job) | Eurobarometer |
|---|---|---|---|
| Racial/ethnic identity {ethnic origin for Eurobarometer) | 0.070 | 0.058 | 0.02 |
| Skin color |  |  | 0.02 |
| Sex | 0.295 | 0.184 | 0.05 |
| Gender identity | 0.070 | 0.042 |  |
| Being transgender |  |  | 0.00 |
| Sexual orientation | 0.019 | 0.015 | 0.01 |
| Disability status | 0.015 | 0.014 | 0.02 |
| Marital status/caregiving responsibilities | 0.127 | 0.088 |  |
| Religion | 0.022 | 0.021 | 0.02 |
| Political views | 0.092 | 0.071 | 0.04 |
| Age | 0.137 | 0.096 | 0.06 |
| Citizenship status | 0.098 | 0.067 |  |
| Place of employment | 0.235 | 0.105 |  |
| Research topics | 0.266 | 0.171 |  |
| Socio-economic situation |  |  | 0.03 |

## 4.3 Comparison with EU Labor Force Survey (Eurostat)

The following is based on the "Labour market situation of migrants and their immediate descendants" (2021) module from the EU-LFS (link). This survey mainly covers the immigrant population; however, some information was collected for all the population 15-74 years old. Of interest for us is a question about the feeling of being discriminated against at work (Table 16). The results refer to all EU countries. They are divided by male and female respondents and the reasons listed are "foreign origin", "gender", "disability", "age", and "other". The overall share of employed women reporting feeling discriminated against at work was 6.1% (vs. 3.6% of men). In this case, the percentages of people reporting that they have been discriminated against based on these factors are much lower than in the EEA, usually between 0.1% and 1%.

Table 30: EEA vs Eurostat

|  | EEA (current job) | | | Eurostat | | |
|---|---|---|---|---|---|---|
|  | All | Male | Female | All | Male | Female |
| In your current job, have you personally experienced discrimination based on: | | | | | | |
| Racial/ethnic identity (EEA) \| Foreign origin (Eurostat) | 0.058 | 0.065 | 0.045 | 0.007 | 0.006 | 0.008 |
| Sex | 0.184 | 0.318 | 0.064 | | | |
| Gender identity | 0.042 | 0.060 | 0.024 | 0.009 | 0.018 | 0.001 |
| Sexual orientation | 0.015 | 0.010 | 0.016 | | | |
| Disability status | 0.014 | 0.020 | 0.009 | 0.001 | 0.001 | 0.001 |
| Marital status / caregiving responsibilities | 0.088 | 0.145 | 0.038 | | | |
| Religion | 0.021 | 0.013 | 0.028 | | | |
| Political views | 0.071 | 0.065 | 0.071 | | | |
| Age | 0.096 | 0.135 | 0.066 | 0.003 | 0.004 | 0.003 |
| Citizenship status | 0.067 | 0.075 | 0.056 | | | |
| Place of employment | 0.105 | 0.113 | 0.101 | | | |
| Research topics | 0.171 | 0.205 | 0.139 | | | |
| Other reason | | | | 0.024 | 0.028 | 0.021 |

## 5. Ethics Statement

The survey questions were prepared in coordination with the VfS Ethics Committee. MinE strictly adhered to EEA's own ethical guidelines ([https://www.eeassoc.org/ethics-and-practices](https://www.eeassoc.org/ethics-and-practices)). Access to the survey response data should be requested to MinE at the first instance, who will relay the request to Dynata. Note that any requests can result in costs as quoted by Dynata, which must be paid for by the requesting party.